\documentstyle[12pt,psfig]{article}

\catcode`\@=11
\long\def\@makefntext#1{
\protect\noindent \hbox to 3.2pt {\hskip-.9pt

$^{{\ninerm\@thefnmark}}$\hfil}#1\hfill}		

\def\@makefnmark{\hbox to 0pt{$^{\@thefnmark}$\hss}}  

\def\ps@myheadings{\let\@mkboth\@gobbletwo
\def\@oddhead{\hbox{}
\rightmark\hfil\ninerm\thepage}
\def\@oddfoot{}\def\@evenhead{\ninerm\thepage\hfil
\leftmark\hbox{}}\def\@evenfoot{}
\def\sectionmark##1{}\def\subsectionmark##1{}}

\renewcommand{\thefootnote}{\fnsymbol{footnote}}

\newcounter{sectionc}\newcounter{subsectionc}\newcounter{subsubsectionc}
\renewcommand{\section}[1] {\vspace*{0.6cm}\addtocounter{sectionc}{1}
\setcounter{subsectionc}{0}\setcounter{subsubsectionc}{0}\noindent
	{\normalsize\bf\thesectionc. #1}\par\vspace*{0.4cm}}
\renewcommand{\subsection}[1] {\vspace*{0.6cm}\addtocounter{subsectionc}{1}
	\setcounter{subsubsectionc}{0}\noindent
	{\normalsize\it\thesectionc.\thesubsectionc. #1}\par\vspace*{0.4cm}}
\renewcommand{\subsubsection}[1]
{\vspace*{0.6cm}\addtocounter{subsubsectionc}{1}
	\noindent {\normalsize\rm\thesectionc.\thesubsectionc.\thesubsubsectionc.
	#1}\par\vspace*{0.4cm}}

\newcounter{appendixc}
\newcounter{subappendixc}[appendixc]
\newcounter{subsubappendixc}[subappendixc]

\renewcommand{\appendix}[1] {\vspace*{0.6cm}
        \refstepcounter{appendixc}
        \setcounter{figure}{0}
        \setcounter{table}{0}
        \setcounter{equation}{0}
        \renewcommand{\thefigure}{\Alph{appendixc}.\arabic{figure}}
        \renewcommand{\thetable}{\Alph{appendixc}.\arabic{table}}
        \renewcommand{\theappendixc}{\Alph{appendixc}}
        \renewcommand{\theequation}{\Alph{appendixc}.\arabic{equation}}
        \noindent{\bf Appendix \theappendixc #1}\par\vspace*{0.4cm}}

\def\abstracts#1{{

\centering{\begin{minipage}{12.2truecm}\footnotesize\baselineskip=12pt\noindent
	\centerline{\footnotesize ABSTRACT}\vspace*{0.3cm}
	\parindent=0pt #1
	\end{minipage}}\par}}

\renewenvironment{thebibliography}[1]
	{\begin{list}{\arabic{enumi}.}
	{\usecounter{enumi}\setlength{\parsep}{0pt}
\setlength{\leftmargin 1.25cm}{\rightmargin 0pt}
	 \setlength{\itemsep}{0pt} \settowidth
	{\labelwidth}{#1.}\sloppy}}{\end{list}}

\newcounter{itemlistc}
\newcounter{romanlistc}
\newcounter{alphlistc}
\newcounter{arabiclistc}

\newcommand{\fcaption}[1]{
        \refstepcounter{figure}
        \setbox\@tempboxa = \hbox{\footnotesize Fig.~\thefigure. #1}
        \ifdim \wd\@tempboxa > 6in
           {\begin{center}
        \parbox{6in}{\footnotesize\baselineskip=12pt Fig.~\thefigure. #1}
            \end{center}}
        \else
             {\begin{center}
             {\footnotesize Fig.~\thefigure. #1}
              \end{center}}
        \fi}

\newcommand{\tcaption}[1]{
        \refstepcounter{table}
        \setbox\@tempboxa = \hbox{\footnotesize Table~\thetable. #1}
        \ifdim \wd\@tempboxa > 6in
           {\begin{center}
        \parbox{6in}{\footnotesize\baselineskip=12pt Table~\thetable. #1}
            \end{center}}
        \else
             {\begin{center}
             {\footnotesize Table~\thetable. #1}
              \end{center}}
        \fi}

\def\@citex[#1]#2{\if@filesw\immediate\write\@auxout
	{\string\citation{#2}}\fi
\def\@citea{}\@cite{\@for\@citeb:=#2\do
	{\@citea\def\@citea{,}\@ifundefined
	{b@\@citeb}{{\bf ?}\@warning
	{Citation `\@citeb' on page \thepage \space undefined}}
	{\csname b@\@citeb\endcsname}}}{#1}}

\newif\if@cghi
\def\cite{\@cghitrue\@ifnextchar [{\@tempswatrue
	\@citex}{\@tempswafalse\@citex[]}}
\def\citelow{\@cghifalse\@ifnextchar [{\@tempswatrue
	\@citex}{\@tempswafalse\@citex[]}}
\def\@cite#1#2{{$\null^{#1}$\if@tempswa\typeout
	{IJCGA warning: optional citation argument
	ignored: `#2'} \fi}}

 1
 1
 1

\font\ninerm=cmr9

\begin{document}

\newcommand{\bmu}{\mbox{\boldmath $\mu$}}
\newcommand{\0}{\,\!}      
\newcommand{\r}[1]{~(\ref{#1})}
\newcommand{\tr}{\mbox{Tr}}
\newcommand{\Dslash}{D\!\!\!\!/}
\newcommand{\jacobian}{{\cal J}\! ac}
\newcommand{\be}{\begin{equation}}
\newcommand{\ee}{\end{equation}}
\newcommand{\bea}{\begin{eqnarray}}
\newcommand{\eea}{\end{eqnarray}}
\newcommand{\A}{{\cal A}}

\begin{flushright}ITP-SB-95-53\end{flushright}
\vspace*{0.5cm}
\centerline{\normalsize\bf Cancellation of quantum mechanical higher loop}
\baselineskip=22pt
\centerline{\normalsize\bf contributions to the gravitational chiral anomaly.}

\vfill
\vspace*{0.6cm}
\centerline{\footnotesize ANDREW K. WALDRON}
\baselineskip=13pt
\centerline{\footnotesize\it Institute for Theoretical Physics,
State University of New York at Stony Brook}
\baselineskip=12pt
\centerline{\footnotesize\it Stony Brook, NY 11794-3840, U.S.A.}
\centerline{\footnotesize E-mail: wally@max.physics.sunysb.edu}
\vspace*{0.3cm}

\vfill
\vspace*{0.9cm}
\abstracts{We give an explicit demonstration, using the
rigorous Feynman rules developed in~$\0^{1}$, that the regularized trace
$\tr \gamma_5 e^{-\beta \Dslash^2}$ for the gravitational chiral anomaly
expressed as an appropriate quantum mechanical path integral is
$\beta$-independent up to two-loop level. Identities and  diagrammatic
notations are developed to facilitate rapid evaluation of graphs given
by these rules.}

\normalsize\baselineskip=15pt
\setcounter{footnote}{0}
\renewcommand{\thefootnote}{\alph{footnote}}

\vspace{.7cm}

It is an old observation of Alvarez-Gaum\'{e} and Witten~\cite{Witten}
that anomalies of quantum field theories, expressed in the
Fujikawa~\cite{Fujikawa} approach as the regulated trace of a jacobian
(${\cal J}$)\be {\cal A}\mbox{(nomaly)}=\lim_{\beta\rightarrow 0}\tr
{\cal J}e^{-\frac{\beta}{\hbar}\hat{R}}\label{anomaly}\ee
may be represented by quantum mechanical path integrals.
However to fully understand such path integrals one must carefully address
issues such as; the precise definition of the measure, which action
corresponds to the particular operator ordering of $\hat{R}$
and the correct Feynman rules for the perturbative expansion of such
path integrals. In their original exposition, Alvarez-Gaum\'{e} and Witten
consider chiral anomalies for which, due to their topological nature,
the expression~(\ref{anomaly}) is $\beta$-independent and calculable
without such subleties via a semiclassical expansion.

Recently, de Boer et.al.~\cite{de Boer} have shown explicity how to
define the measure, action and Feynman rules for quantum mechanical
path integrals for both bosons and fermions in curved space. The exact rules
they obtain, although novel, follow directly from a rigorous treatment of the
measure constructed from insertions of complete sets of coherent states.
In this note we begin with their results and return to the gravitational
chiral anomaly to verify through two loop order that their Feynman rules
give the correct \underline{$\beta$-independent} result for~(\ref{anomaly}).

The gravitational chiral anomaly is given by the index of the Dirac operator
\be {\cal A} = \tr \gamma_5 e^{-\frac{\beta}{\hbar}\Dslash\Dslash}
=n_+-n_-\ee where $n_{\pm}$ are the number of positive/negative parity
zero modes of $\Dslash=e_a\0^\mu\gamma^a(\partial_\mu+
\frac{1}{4}\omega_\mu\0^{ab}\gamma_a\gamma_b)$. ${\cal A}$ may be represented
by a quantum mechanical path integral
\be \A=\int_M\frac{d^ny\sqrt{g(y)}}{(2\pi i)^2}d^n\Psi \ Z[y^\mu,\Psi^a],
\label{anom}\ee
where the path integral $Z$ depends on constant
background fields $y^\mu$ and real
fermionic (Majorana) $\Psi^a$ ($a=1\ldots n=\mbox{dim} M$
for some $n$-manifold $M$).
Schematically \be Z[y^\mu,\Psi^a]=
\int[dq^\mu(t)d\psi^a(t)da^\mu(t)db^\mu(t)dc^\mu(t)]
\exp\{-\frac{1}{\hbar}(S_{\em kin}+S_{\em int})\},\ee
where\footnote{We make two minor deviations
from the notation of~\cite{de Boer},
firstly the worldline parameter $t\in[0,1]$ not $[-1,0]$ and the
anticommuting ghosts appear
as $g_{\mu\nu}b^\mu c^\nu$ rather than $(1/2)g_{\mu\nu}b^\mu c^\nu$.}
\bea S_{\rm kin}&=&\int_{0}^{1}dt[\bar{\chi}_A\dot{\chi}^A
+\frac{1}{2}g_{\mu\nu}(y)(\dot{q}^\mu\dot{q}^\nu+
a^\mu a^\nu+2b^\mu c^\nu)]\\
S_{\rm int}&=&\int_{0}^{1}dt\frac{1}{2}\left[\left(
g_{\mu\nu}(q\!+\!y)-g_{\mu\nu}(y)\right)
\left(\dot{q}^\mu\dot{q}^\nu+a^\mu a^\nu+2b^\mu c^\nu
\right)\right.
\nonumber\\ &&\left.
+\omega_{\mu ab}(q\!+\! y)\dot{q}^\mu(\Psi+\psi)^a(\Psi+\psi)^b
+\frac{\hbar^2}{4}\left(\Gamma^\alpha_{\beta\mu}\Gamma^\beta_{\alpha \nu}
+\frac{1}{2}\omega_{\mu ab}\omega_\nu\0^{ab}\right) (q\!+\!y)
g^{\mu \nu}(q\!+\!y)\right] .\eea
As emphasized in~\cite{de Boer}, the above expression
is the continuum limit of a rigorous dicrete result for $Z[y^\mu,\Psi^a]$.
Propagators must be derived from the discrete expressions and in this way
ambiguities arising from products of distributions are resolved. Vertices
may, however, be read directly from the continuum $S_{\em int}$ given
above. In the kinetic action we have written $n/2$ complexified spinors
$\chi^A$ as reminder that the correct Majorana propagators are obtained
using the complexified $\chi^A$ as an intermediate step in order to construct
fermionic coherent states. One could, of course, work completely in the
complex basis but the interactions are then more complicated and depend on
the arbitrary choice of complexification.
The terms proportional to $\hbar^2$ in $S_{\rm int}$ occur since the
path integral was derived by Weyl ordering the regulator
$\Dslash\Dslash$ so that matrix elements could be calculated using
the midpoint rule. However they contribute as extra vertices at three
and higher loop level only so can be disregarded here. Of course,
in higher loops we do expect these extra vertices to conspire with the new
Feynman rules to give a $\beta$ independent result.

The number of graphs we must consider is greatly reduced if we choose
Riemann normal coordinates $y^\mu$ such that geodesics through $O$
(say) are straight lines. I.e. any geodesic through $O$ with arc length $s$
has coordinates $y^\mu(s)=\xi^\mu s$. Taylor expanding in $s$ gives
$y^\mu(s)=y^\mu(O)+s\dot{y}^\mu(O)+
(1/2!)s^2\ddot{y}^\mu(O)+\cdots$ from
which we see that the second and higher derivatives of $y^\mu(s)$
at $O$ vanish. But the geodesic equation
$\ddot{y}^\mu(s)+\Gamma^\mu_{\alpha\beta}\dot{y}^\alpha(s)\dot{y}^\beta(s)=0$
evaluated at $O$ gives $\Gamma^\mu_{\alpha\beta}(O)=0$ and taking
higher derivatives w.r.t. $s$ about $O$ of the geodesic equation shows that
all symmetrized derivatives of the connection coefficients vanish
$\partial^n_{(\alpha_1\cdots\alpha_n}\Gamma^\mu_{\alpha\beta)}(O)=0$.
It is now easy to find a covariant expression for derivatives on any
tensor at $O$, in particular for the metric one finds
\bea
\partial_\alpha g_{\mu\nu}(O)&=&0\\
\partial^2_{\alpha_1\alpha_2} g_{\mu\nu}(O)&=&
\frac{1}{3}R_{\mu(\alpha_1\alpha_2)\nu}.
\eea
We also need Riemann normal coordinates for spinors to handle derivatives
on the spin connection. Let us choose frames $e_\mu\0^a$ such that the
components
of any vector with a flattened index $v^a(s)$ parallely transported along
a geodesic through $O$ are constant $\dot{v}^a(s)=0$. Taylor expanding
$v^a(s)=v^a(O)+s\dot{v}^a(O)+(1/2!)s^2\ddot{v}^a(O)+\cdots$ shows
that all derivatives of $v^a(s)$ vanish at $O$ so that the parallel
transport equation $Dv^a(s)/Ds=\dot{v}^a(s)+
\dot{y}^\mu\omega_\mu\0^a\0_bv^b(s)=0$
and derivatives thereof give that all symmetrized derivatives of the
spin connection vanish
$\partial^n_{(\alpha_1\cdots\alpha_n}\omega_{\mu)}\,\! ^a \,\! _b(O)=0$
for $n=0,1,\ldots$. Hence we readily obtain
\bea
\omega_\mu\0^a\0_b(O)&=&0\\
\partial_\alpha\omega_\mu\0^a\0_b(O)&=&\frac{1}{2}R_{\alpha\mu}\0^a\0_b\\
\partial^2_{\alpha_1\alpha_2}\omega_\mu\0^a\0_b(O)&=&
\frac{1}{3}D_{(\alpha_1}R_{\beta)\mu}\0^a\0_b.
\eea
Let us now give all vertices relevant  to our calculations.
\be
\begin{array}{ll}
\raisebox{-1cm}{\psfig{figure=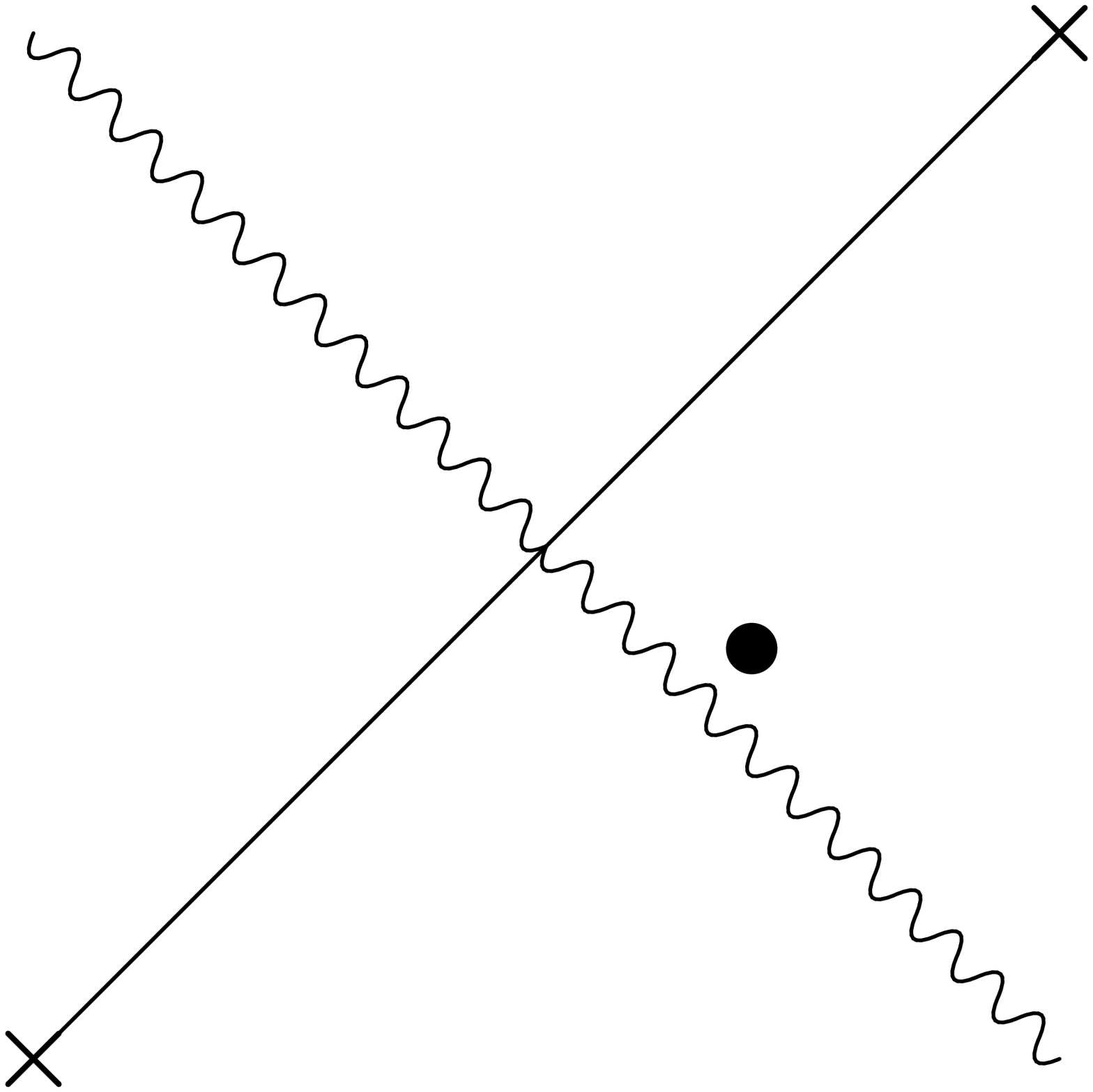,width=2cm}}
=-\frac{1}{4}R_{\mu\nu ab}\Psi^a\Psi^b\int_{0}^{1}q^\mu\dot{q}^\nu &
\raisebox{-1cm}{\psfig{figure=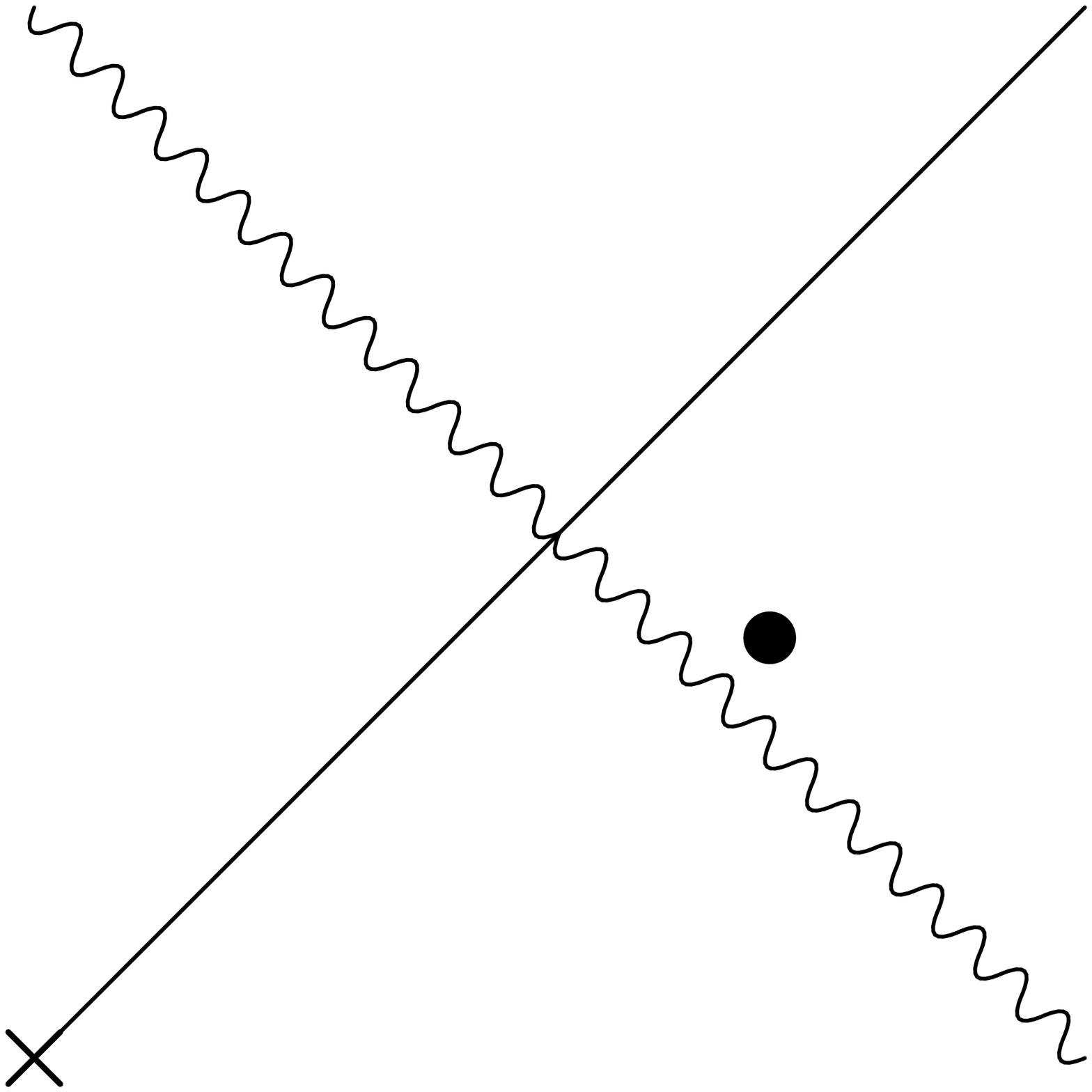,width=2cm}}
=\frac{1}{2}R_{\mu\nu ab}\Psi^b\int_{0}^{1}q^\mu\dot{q}^\nu\psi^a\\&\\
\raisebox{-1cm}{\psfig{figure=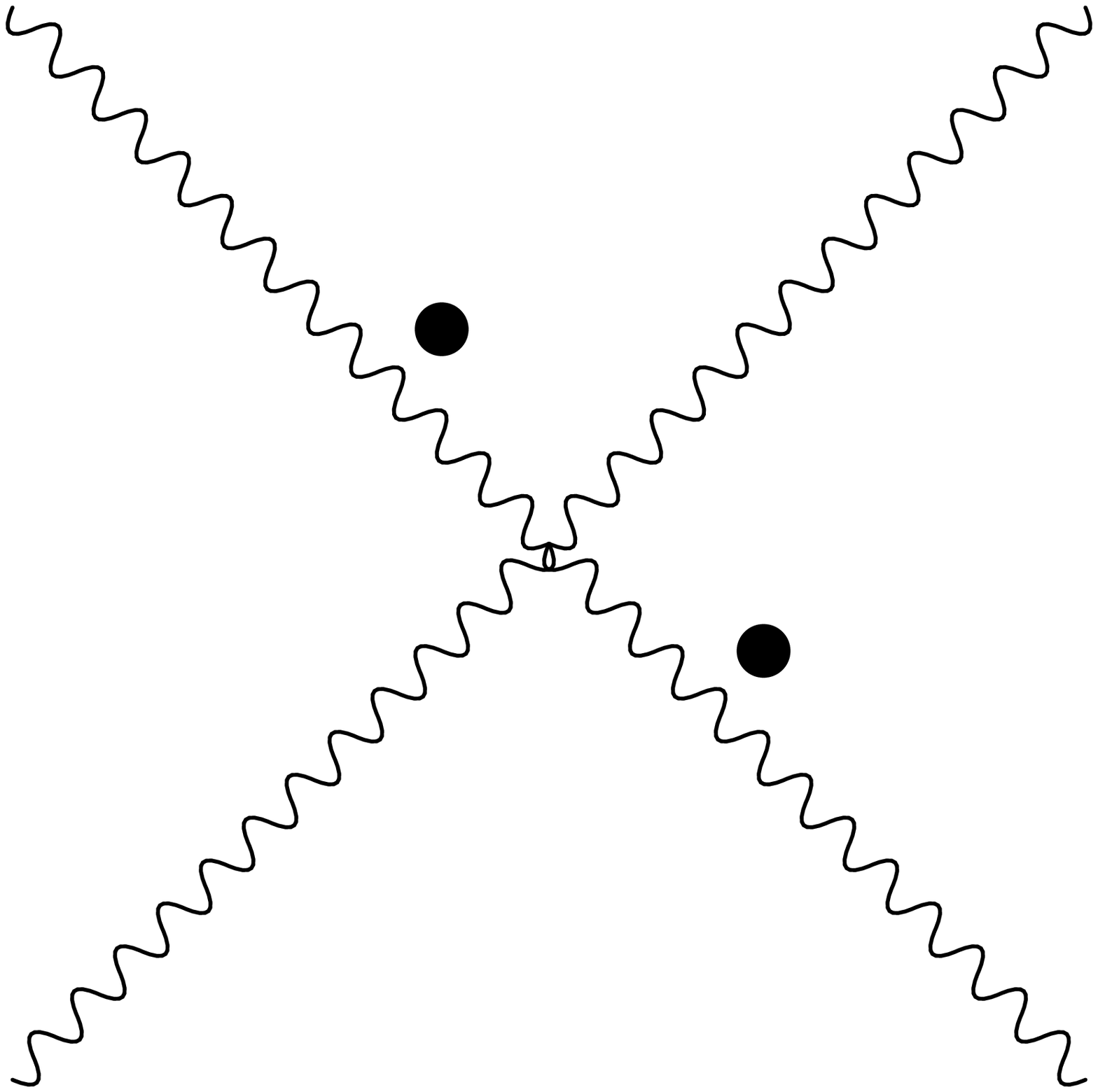,width=2cm}}
=-\frac{1}{3}R_{\mu(\alpha\beta)\nu}\frac{1}{2!2!}
\int_{0}^{1}q^\mu q^\nu \dot{q}^\alpha\dot{q}^\beta &
\raisebox{-1cm}{\psfig{figure=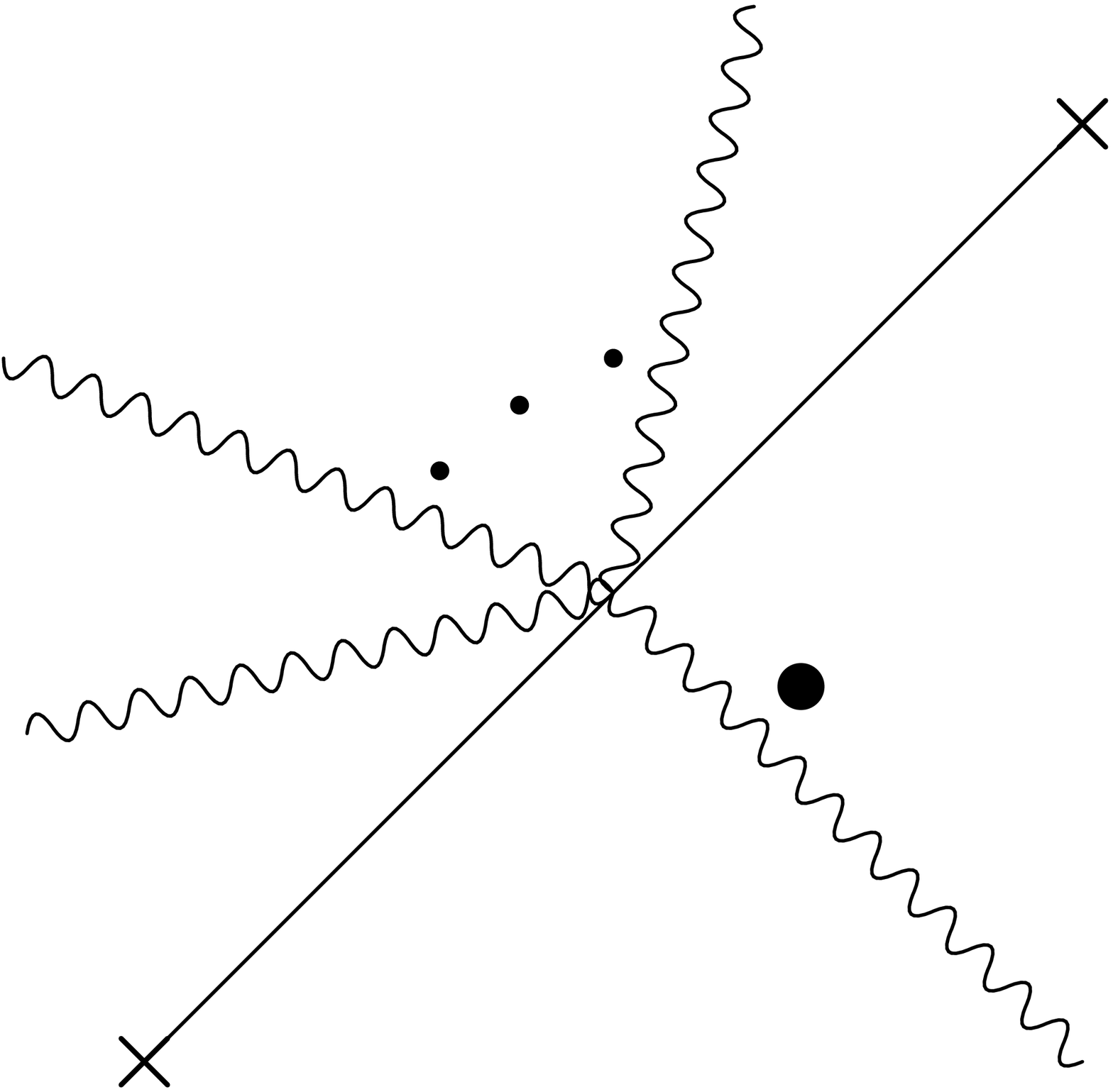,width=2cm}}
=-\frac{1}{2}\partial^n_{\alpha_1\cdots\alpha_n}\omega_{\mu ab}\Psi^a\Psi^b
\frac{1}{n!}\int_{0}^{1}q^{\alpha_1}\cdots q^{\alpha_n} \dot{q}^\mu
\end{array}\label{vertices}
\ee
Wiggly lines denote bosons and straight lines Majorana fermions.
A dot on a boson line indicates a $\dot{q}^\mu$ at the vertex
and the on the end of fermion lines denote external fermion
background fields $\Psi^a$. It is expedient to leave the factor
$\partial^n_{\alpha_1\cdots\alpha_n}\omega_{\mu ab}$ as it stands rather than
grinding out directly its covariant Riemann normal coordinate expression
since in graphs only certain antisymmetrized combinations will appear
which are then readily covariantized. Note that we ignore ghost vertices.
Strictly speaking we should replace $\frac{1}{2!2!}
\int_{0}^{1}q^\mu q^\nu \dot{q}^\alpha\dot{q}^\beta\rightarrow
\frac{1}{2!}\int_{0}^{1}q^\mu q^\nu (\frac{1}{2!}\dot{q}^\alpha\dot{q}^\beta
+\frac{1}{2!}a^\alpha a^\beta+b^\alpha c^\beta)$, however at two loop level
the ghosts only arise in self energy loops where they exactly cancel the
delta function divergence of the accompanying $\dot{q}\dot{q}$ self energy
contraction.

We now give the propagators of the theory,
\bea
\langle q^\mu(s) q^\nu(t)\rangle&=&g^{\mu\nu}(y)\Delta(s,t)\\
\langle q^\mu(s) \dot{q}^\nu(t)\rangle&=&g^{\mu\nu}(y)\Delta\!^\cdot(s,t)\\
\langle \dot{q}^\mu(s) \dot{q}^\nu(t)\rangle&=&
g^{\mu\nu}(y)^\cdot\!\Delta\!^\cdot(s,t)\\
\langle a^\mu(s) a^\nu(t)\rangle&=&
g^{\mu\nu}(y)(^\cdot\!\Delta\!^\cdot(s,t)+1)\\
\langle b^\mu(s) c^\nu(t)\rangle&=&
-g^{\mu\nu}(y)(^\cdot\!\Delta\!^\cdot(s,t)+1)\\
\langle\psi^a(s)\psi^b(t)\rangle&=&
\delta^{ab}(1/2)\epsilon(s-t)+K^{ab}.
\eea
Where we have denoted
\bea
\Delta(s,t)\equiv&\raisebox{-.2cm}{\psfig{figure=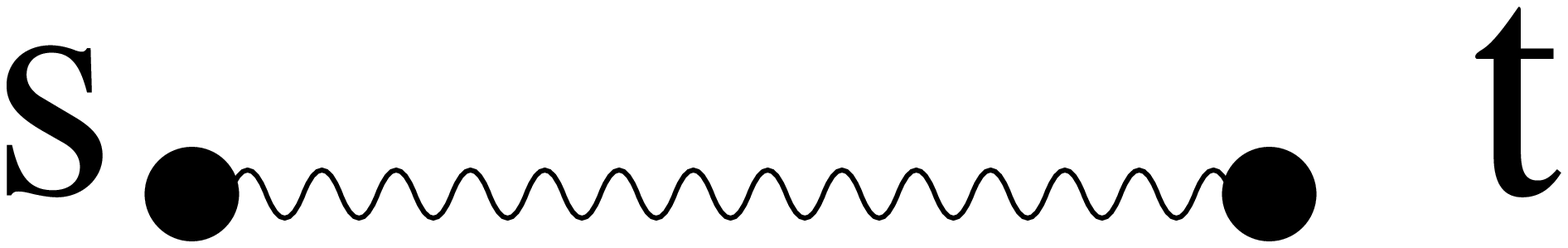,width=2cm}}&
=t(1-s)\theta(s-t)+s(1-t)\theta(t-s)\label{prop1}\\
\Delta\!^\cdot(s,t)\equiv&
\raisebox{-.2cm}{\psfig{figure=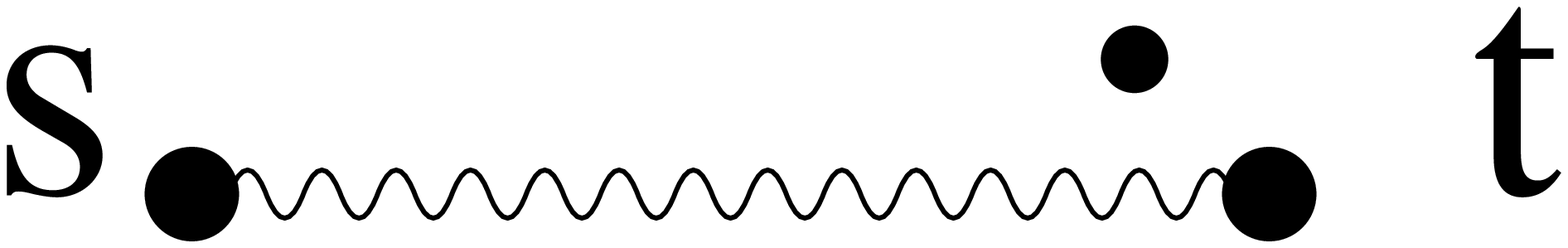,width=2cm}}&
=(1-s)\theta(s-t)-s\theta(t-s)=(d/dt)\Delta(s,t)\\
^\cdot\!\Delta\!^\cdot(s,t)\equiv&
\raisebox{-.2cm}{\psfig{figure=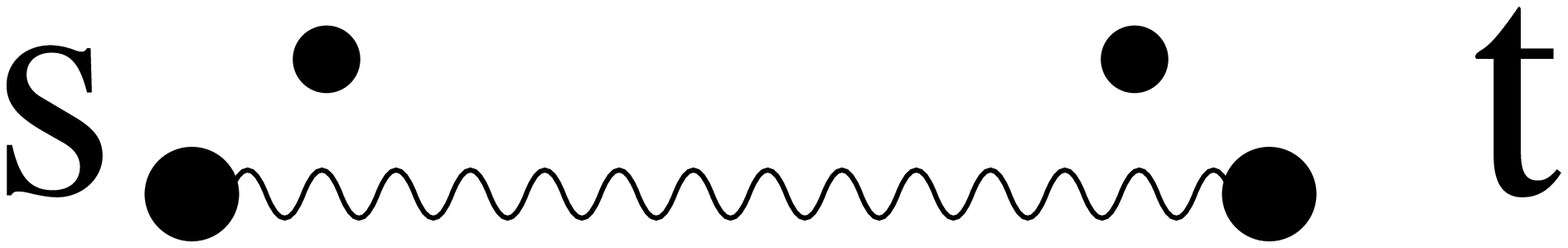,width=2cm}}&
=\delta(s-t)-1=(d^2/dsdt)\Delta(s,t).\label{prop3}
\eea
Further the theta function is defined as
\be\theta(s-t)\equiv\left\{ \begin{array}{lr}1&\hspace{.5cm} s>t\\
\frac{1}{2}&s=t\\0 & s<t
\end{array}\right. \ee
and $\epsilon(s-t)\equiv\theta(s-t)-\theta(t-s)$. When handling products of
distributions the delta function is treated as a Kronecker delta, for example
$\int_{0}^{1}ds\theta(s-t)\delta(s-t)=1/2$. The term $K^{ab}$ in the fermion
propagator is a relic of the complexification (see~\cite{de Boer})
 of the original real spinors,
here we need only that $K^{ab}=-K^{ab}$. Of course final (physical)
results should be $K^{ab}$ independent.

Now all graphs are of the factorized form $(\mbox{general rel.})
\times(\mbox{integrals over }\Delta,
\Delta\!^\cdot,\ ^\cdot\!\Delta\!^\cdot
\mbox{ and } \epsilon)$, which are, in principle, elementary to perform.
Of course in practice there is a large amount of trivial algebra to perform
which can be greatly simplified if one first derives certain identities
for the propagators~(\ref{prop1})-(\ref{prop3}).
Firstly observe that propagators of the form
${\cal D}(s,t)=d_1(s,t)\theta(s-t)+d_2(s,t)\theta(t-s)$ close under
multiplication
\bea
[{\cal D}\tilde{\cal D}](s,t)&\equiv&
\int_{0}^{1}dr{\cal D}(s,r){\tilde{\cal D}}(r,t)\nonumber\\
&=&\theta(s-t)\left(\int_{0}^{t}d_1(s,r)\tilde{d}_2(r,t)
+\int_{t}^{s}d_1(s,r)\tilde{d}_1(r,t)
+\int_{s}^{1}d_2(s,r)\tilde{d}_1(r,t)\right)\nonumber\\
&+&\theta(t-s)\left(\int_{0}^{s}d_1(s,r)\tilde{d}_2(r,t)
+\int_{s}^{t}d_2(s,r)\tilde{d}_2(r,t)
+\int_{t}^{1}d_2(s,r)\tilde{d}_1(r,t)\right).\label{multiply}
\eea
Let us now adopt a diagrammatic notation in which propagators are depicted as
in~(\ref{prop1})-(\ref{prop3}) where a dot at the end of a line or
vertex denotes a point yet to be
integrated over (see the ends of the propagators above, we usually also
attach a variable $s,t,\ldots$ for clarity) and a cross denotes a point
where an integration $\int_{0}^{1}$ has been performed. For example,
applying~(\ref{multiply}) in this notation
\bea
\raisebox{-.2cm}{\psfig{figure=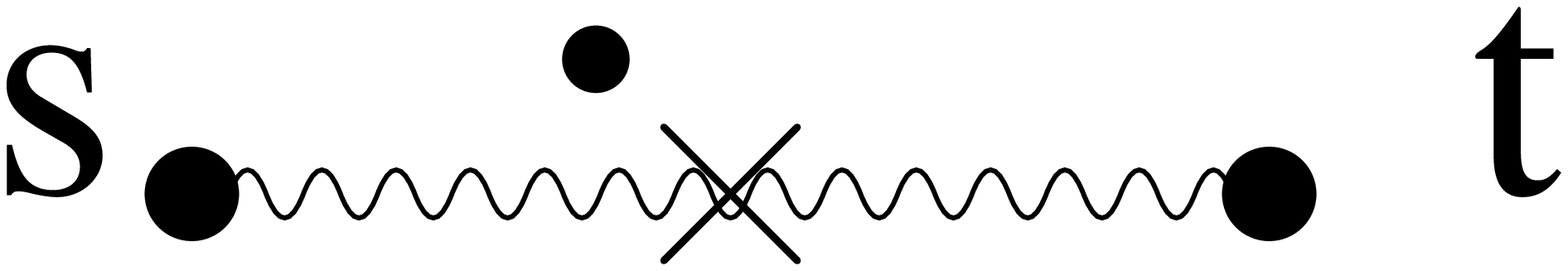,width=2cm}}
&\equiv&\int_{0}^{1}dr\Delta\!^\cdot(s,r)\Delta(r,t)\nonumber \\
&=&(1/2)(s-t)(t(1-s)\theta(s-t)+s(1-t)\theta(t-s))\nonumber\\
&=&(1/2)(s-t)\Delta(s,t)\nonumber\\
&=&-\left(\raisebox{-.2cm}{\psfig{figure=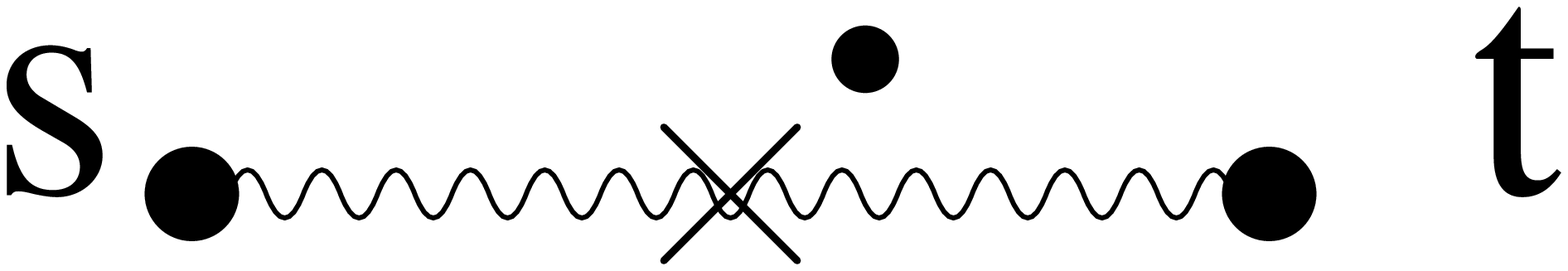,width=2cm}}
\right)\label{elvis}
\eea
The last line of~(\ref{elvis}) is an example of an allowed (and very
useful) integration by parts. In~\cite{de Boer} it is stressed that
ad hoc integrations by parts are not compatible with the Kronecker delta
prescription for the delta function, however~(\ref{elvis}) is correct since
it is an example of the more general result
\be
(d^n/dt^n)\Delta(0,t)=0=(d^n/dt^n)\Delta(1,t).
\ee
It is highly expedient to make integrations by parts at vertices with
a single dotted line, in diagrammatic notation\footnote{Here we assume that
the graph is expressed only in terms of propagators $\Delta$,
$\Delta\!^\cdot$ and $^\cdot\!\Delta\!^\cdot$.}
\be
\raisebox{-1.3cm}{\psfig{figure=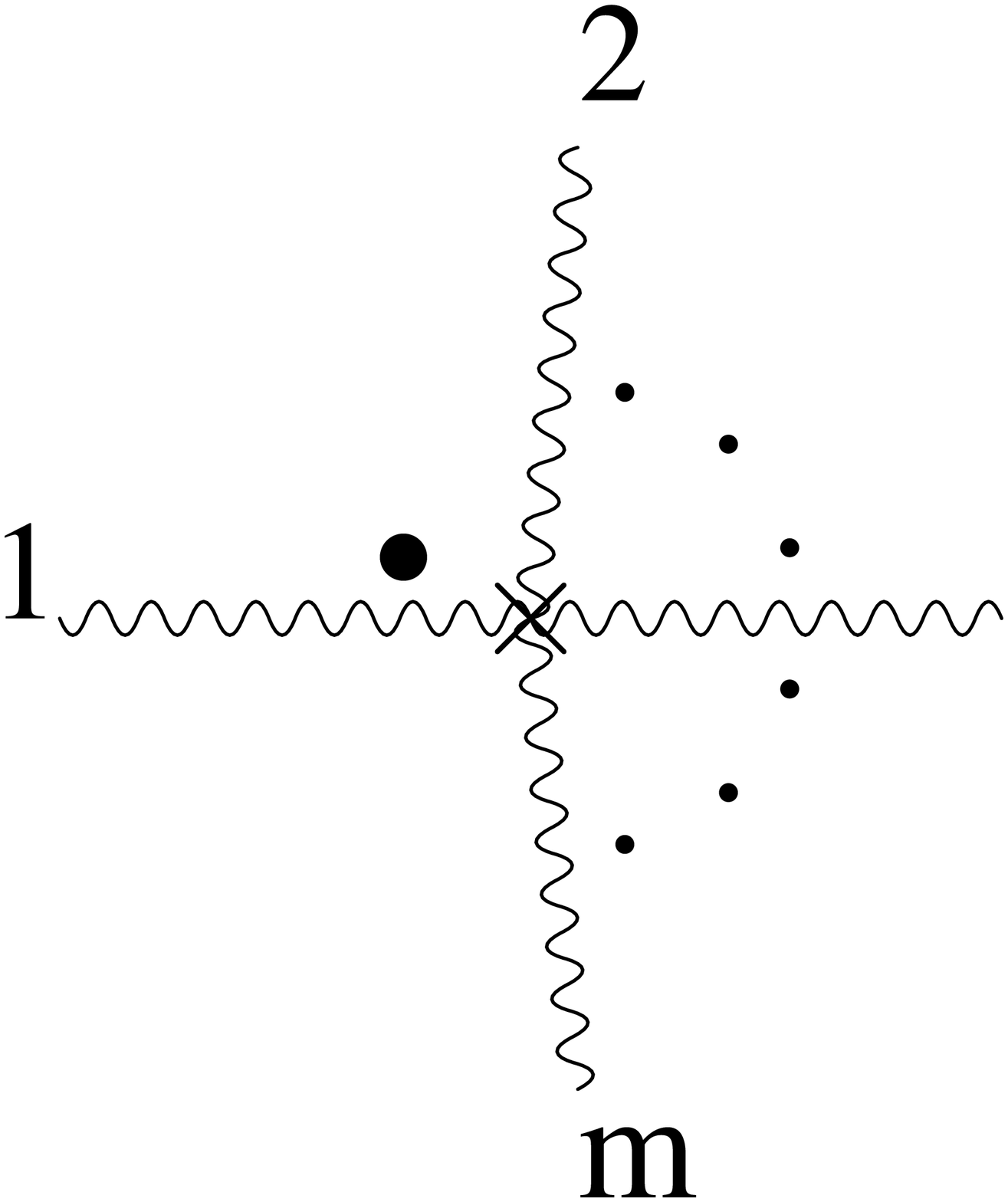,width=2.4cm}}=-\sum_{i=2}^{m}
\raisebox{-1.3cm}{\psfig{figure=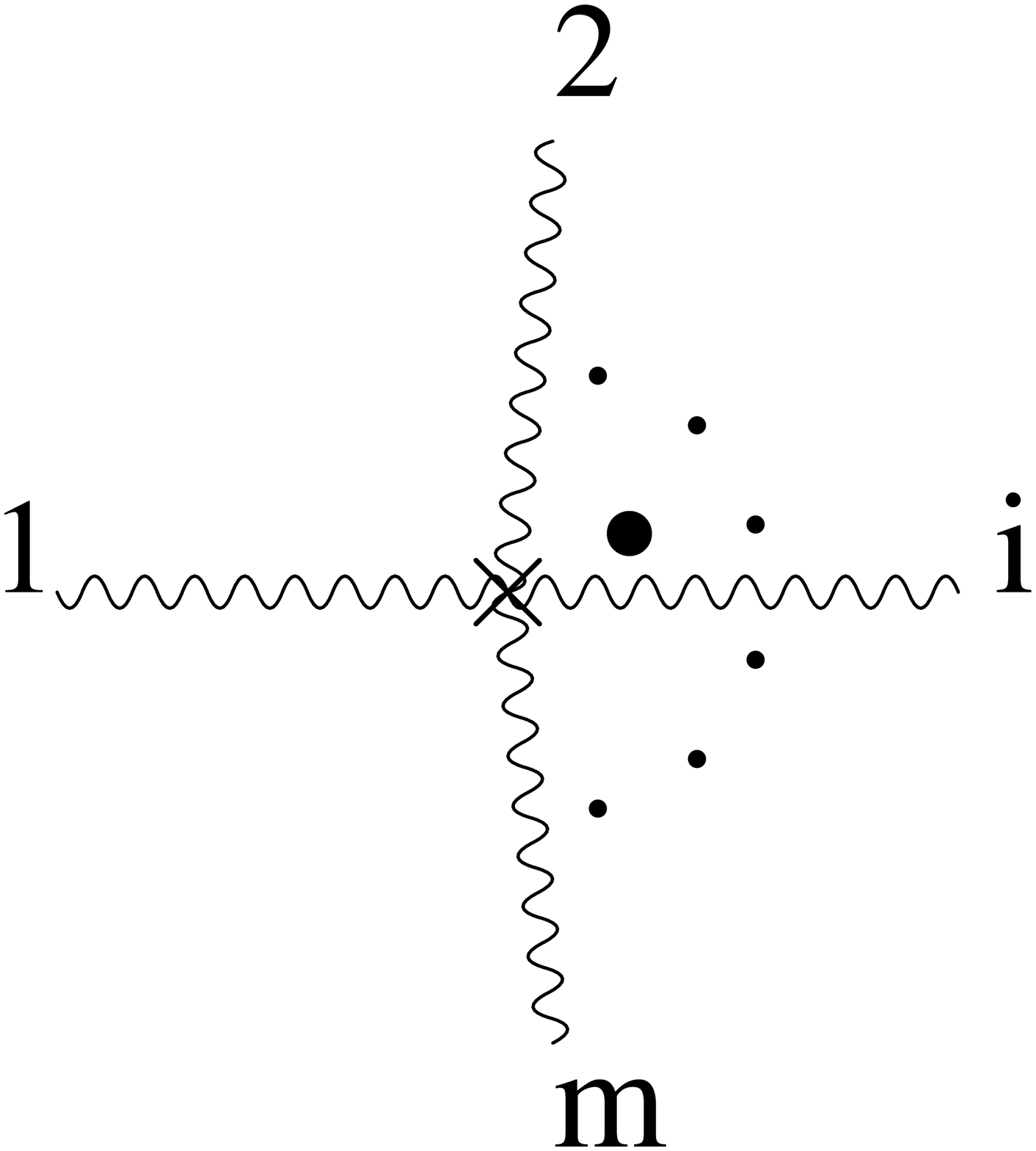,width=2.4cm}}\ .
\ee
This relation holds even if the outgoing lines form self-energy loops
since $\Delta\!^\cdot(s,s)=(1/2)(d/ds)$ $\Delta(s,s)$.

The propagator $^\cdot\!\Delta\!^\cdot(s,t)$
seems to be an odd bunny but may be easily handled by noticing that its
effect in any graph is to produce the difference of two graphs, the
first in which the points $s$ and $t$ are pinched together
(doing the delta function) and the second in
 which the $^\cdot\!\Delta\!^\cdot$
is simply absent (see the example in figure 1.).
The only exception (at two loops, although a similar
statement holds at higher loops) are self energy loops where,
due to the aforementioned ghosts, $\Delta(s,s)=\delta(0)-1\longrightarrow-1$.

In table 1 we list the results for various products of propagators,
plus the results of integrating over the ends of these concatenated
propagators and forming loops from them, included also are the results
for the graphs with
fermion propagators in which we denote
$(1/2)\epsilon(s-t)\equiv
\raisebox{-.15cm}{\psfig{figure=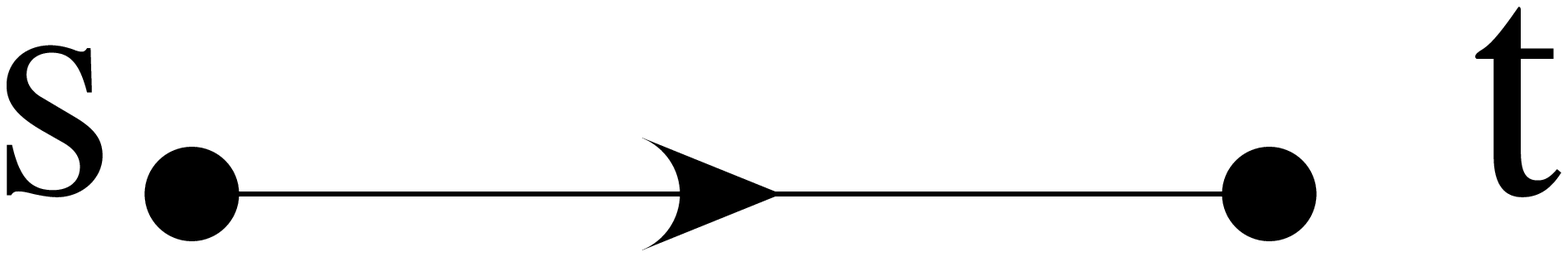,width=2cm}}$.
Such graphs may also easily be evaluated using the techniques discussed
above if one makes use of the identity
\be\raisebox{-.15cm}{\psfig{figure=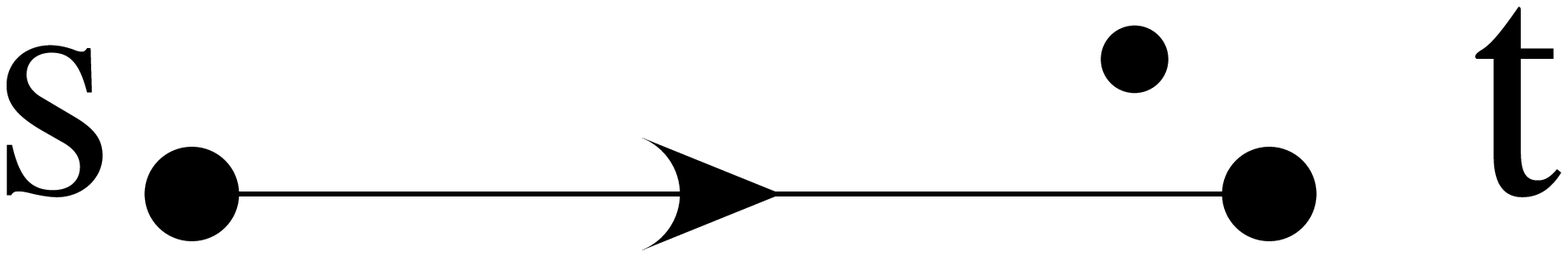,width=2cm}}=
\frac{1}{2}\frac{d}{dt}\epsilon(s-t)=-\delta(s-t),\ee
for convenience however, we give the explicit results for these graphs.
To avoid confusion, note that table 1 is really just a table of integrals if
one decodes the graphical notation used here.

\begin{figure}
$$ \begin{array}{|c|}
\hline \\ \raisebox{-.8cm}{\psfig{figure=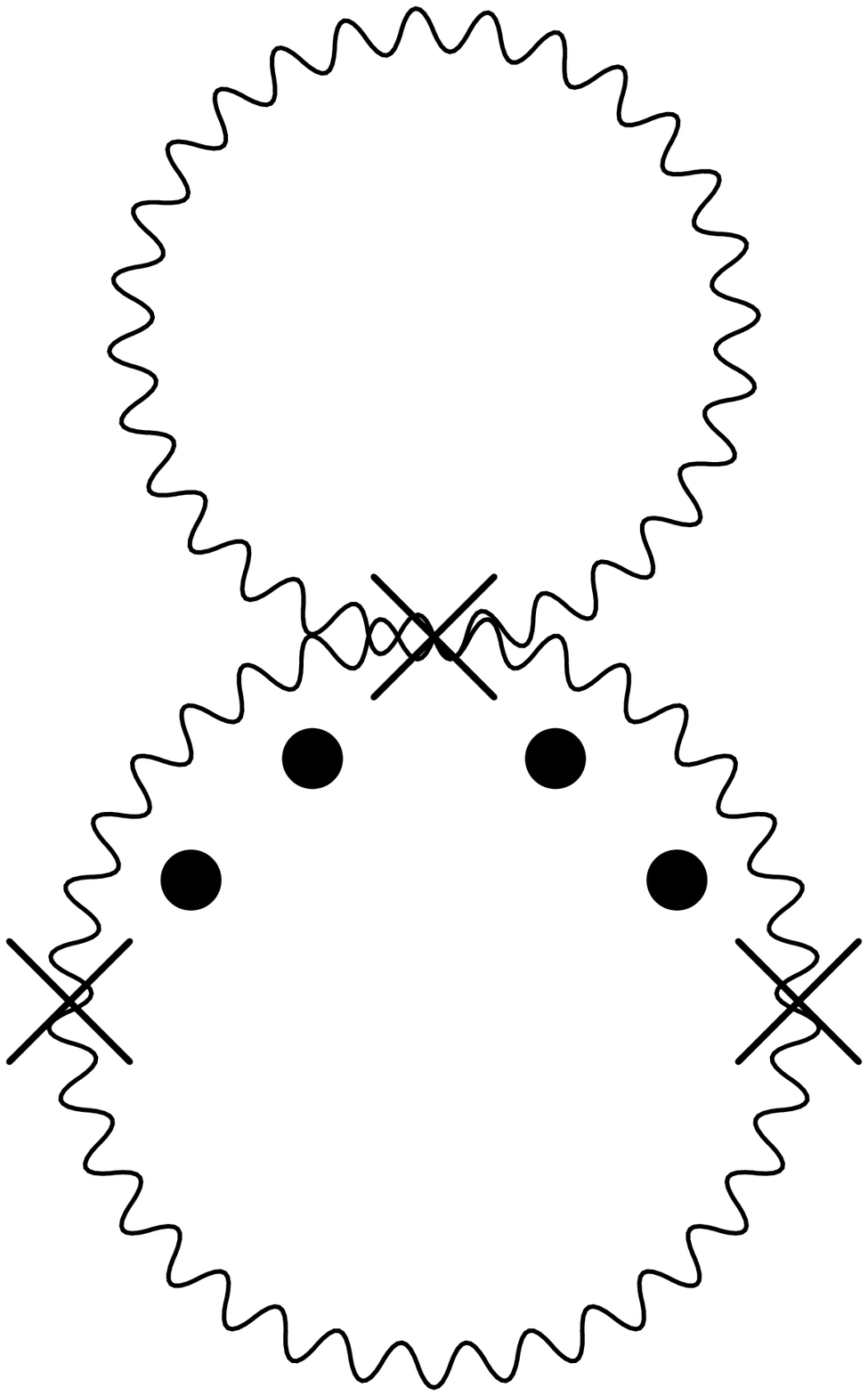,width=1.3cm}}\ \ =\ \
\raisebox{-.7cm}{\psfig{figure=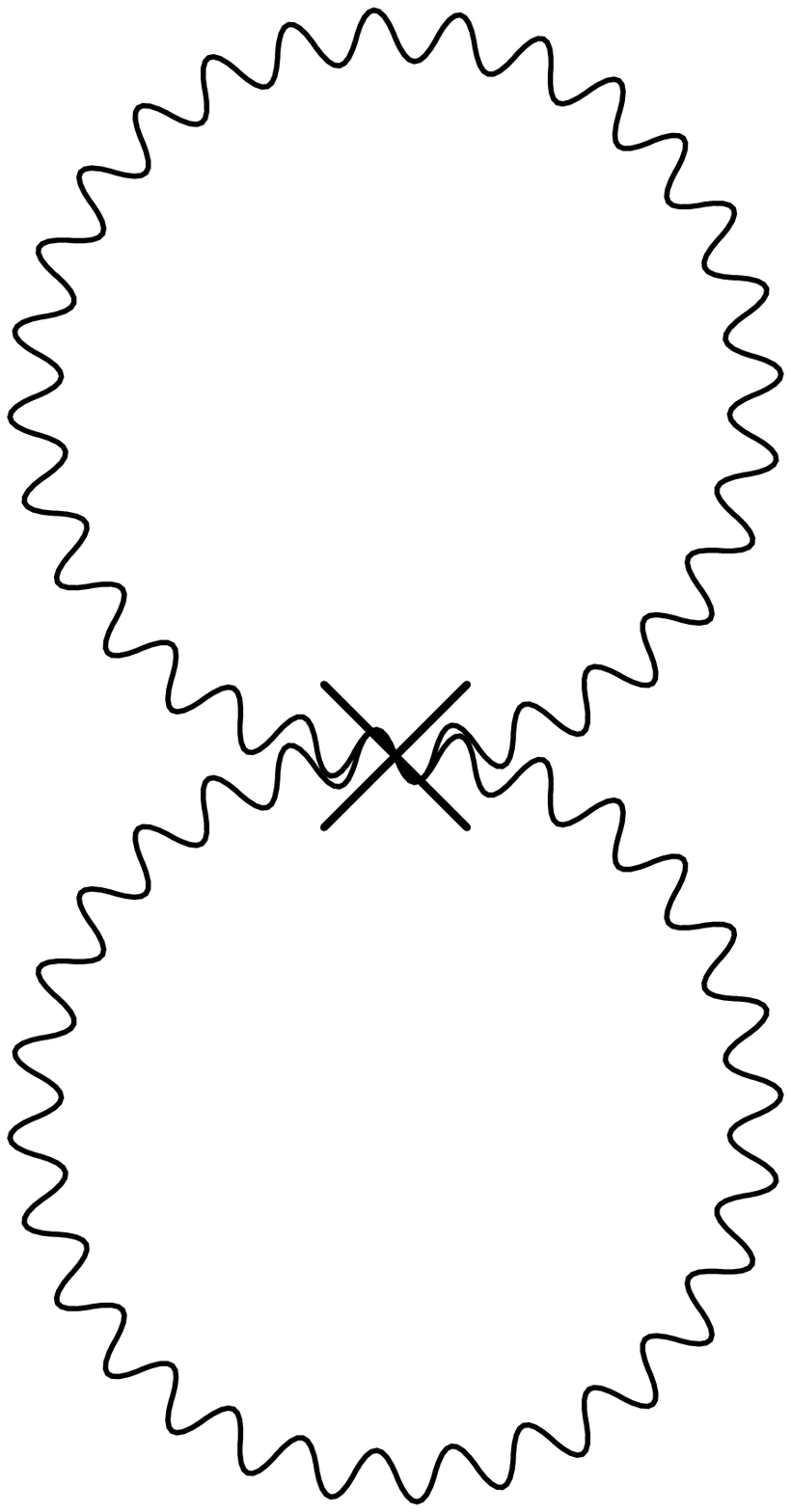,width=1cm}}
\ \ \ -2\ \ \raisebox{-.7cm}{\psfig{figure=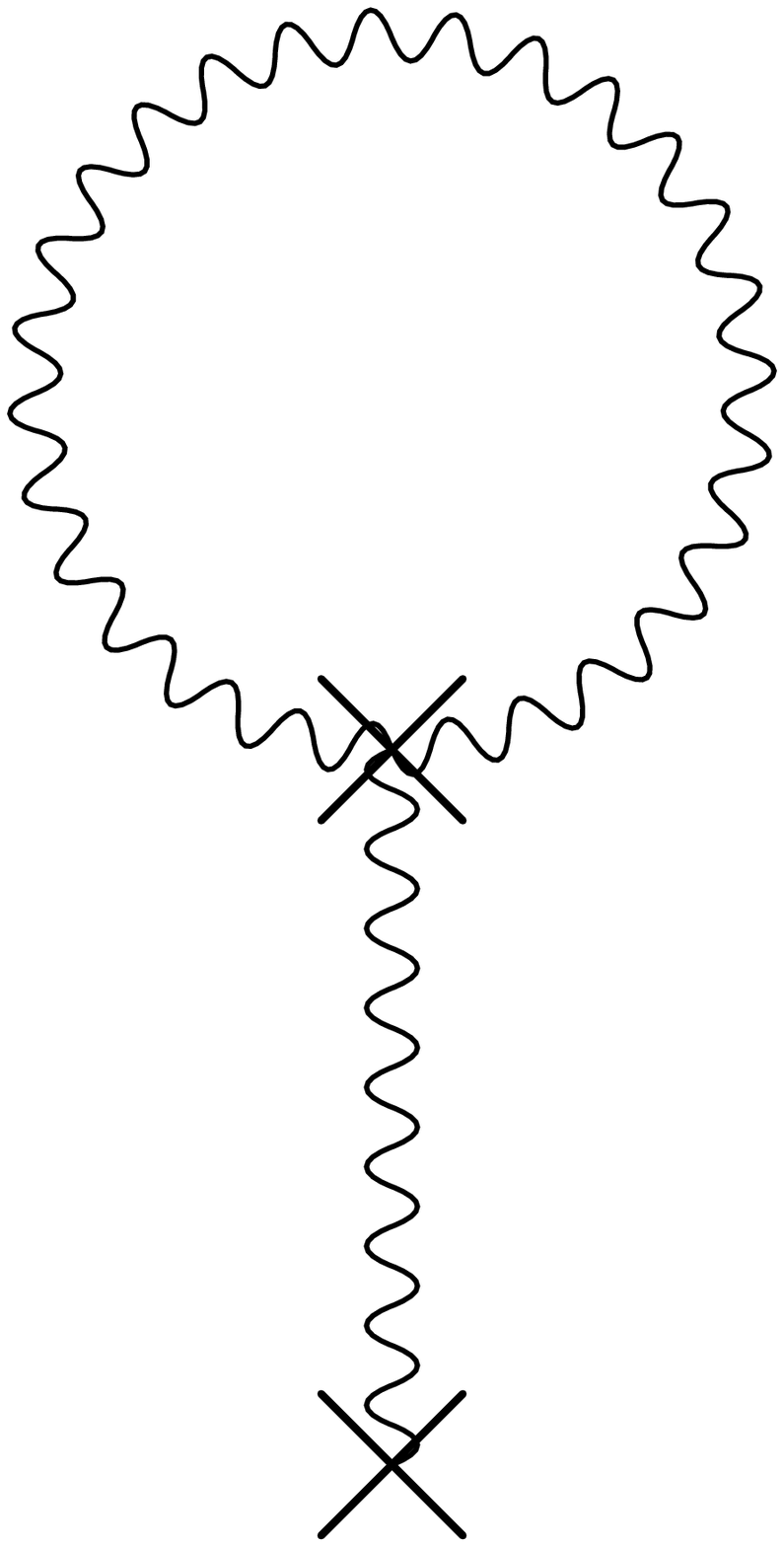,width=1cm}}
\ \ +\ \ \raisebox{-.5cm}{\psfig{figure=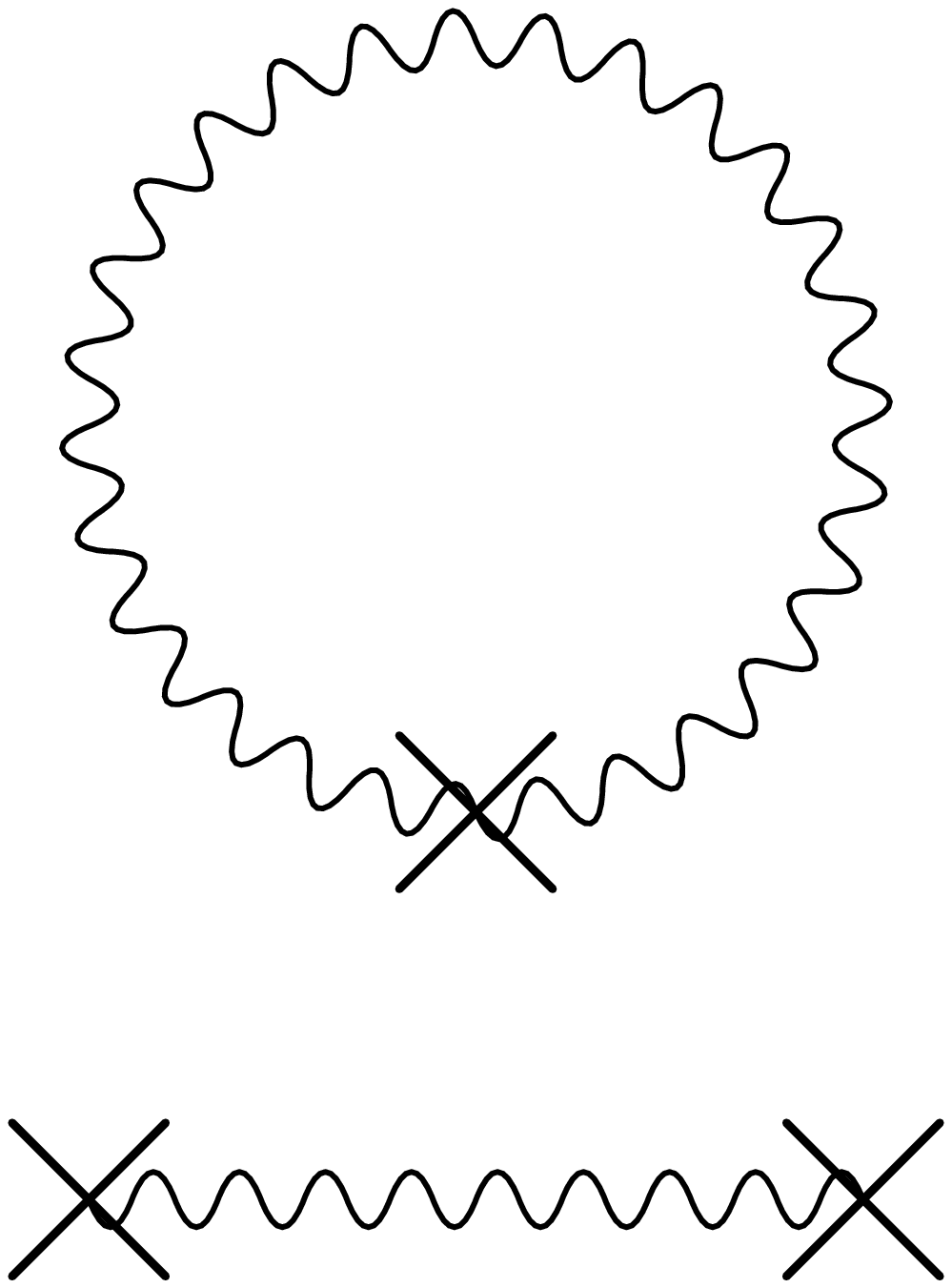,width=1.2cm}}
\\ \\
\raisebox{-.5cm}{\psfig{figure=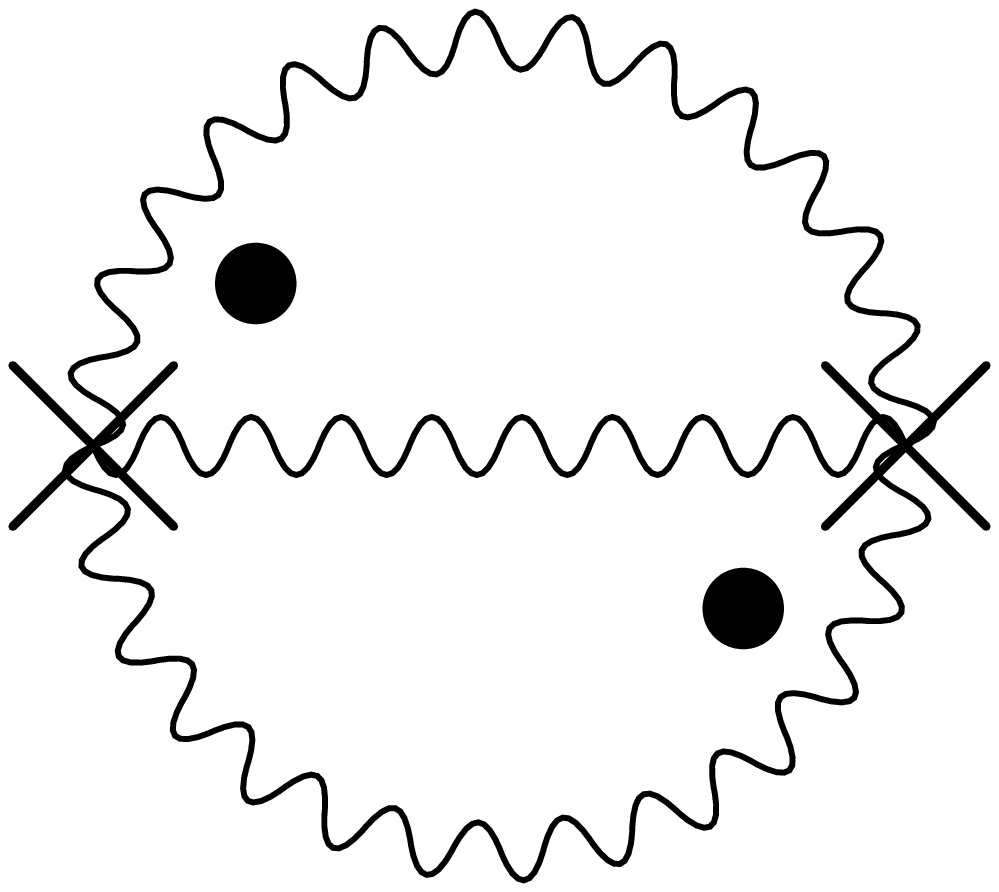,width=1.3cm}}\ \ =\ \
-\frac{1}{2}\ \ \raisebox{-.5cm}{\psfig{figure=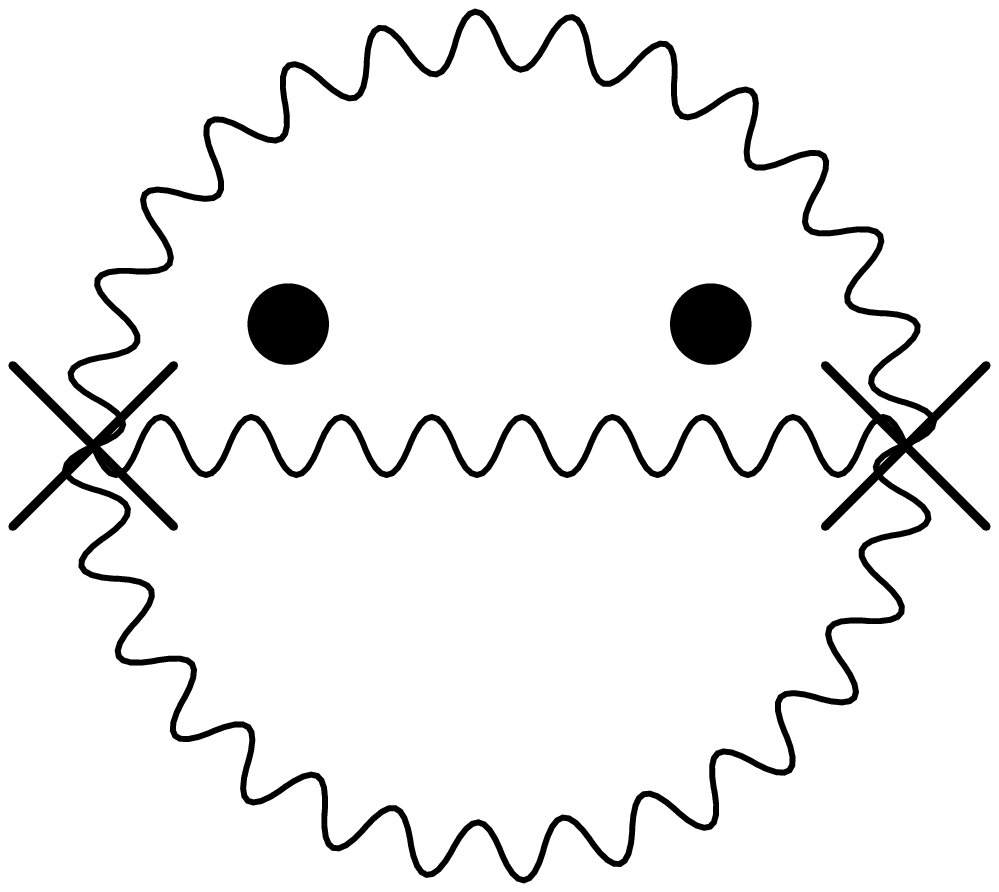,width=1.3cm}}
\ \ = \ \
-\frac{1}{2}\ \ \raisebox{-1.1cm}{\psfig{figure=diagram11.ps,width=1.2cm}}
\ \ + \ \
\frac{1}{2}\ \ \raisebox{-.5cm}{\psfig{figure=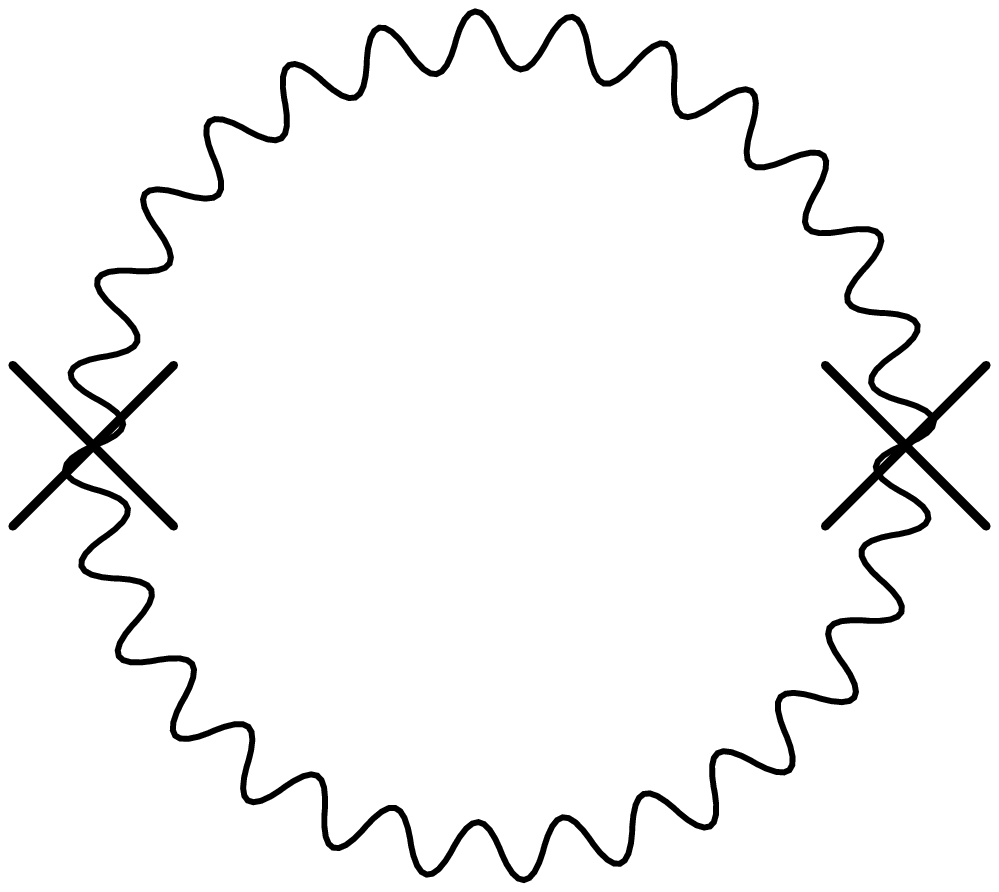,width=1.3cm}}
\\ \\
\hspace{.7cm}
\raisebox{-.3cm}{\psfig{figure=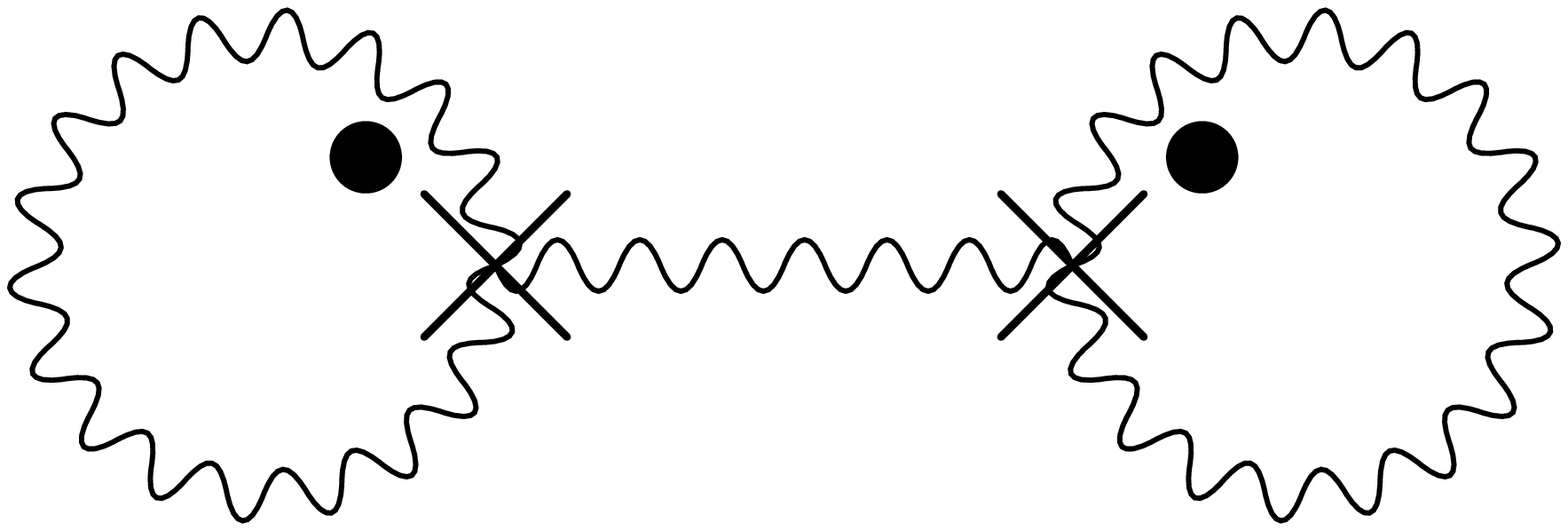,width=2cm}}\ \ =\ \
-\frac{1}{2}\ \ \raisebox{-.3cm}{\psfig{figure=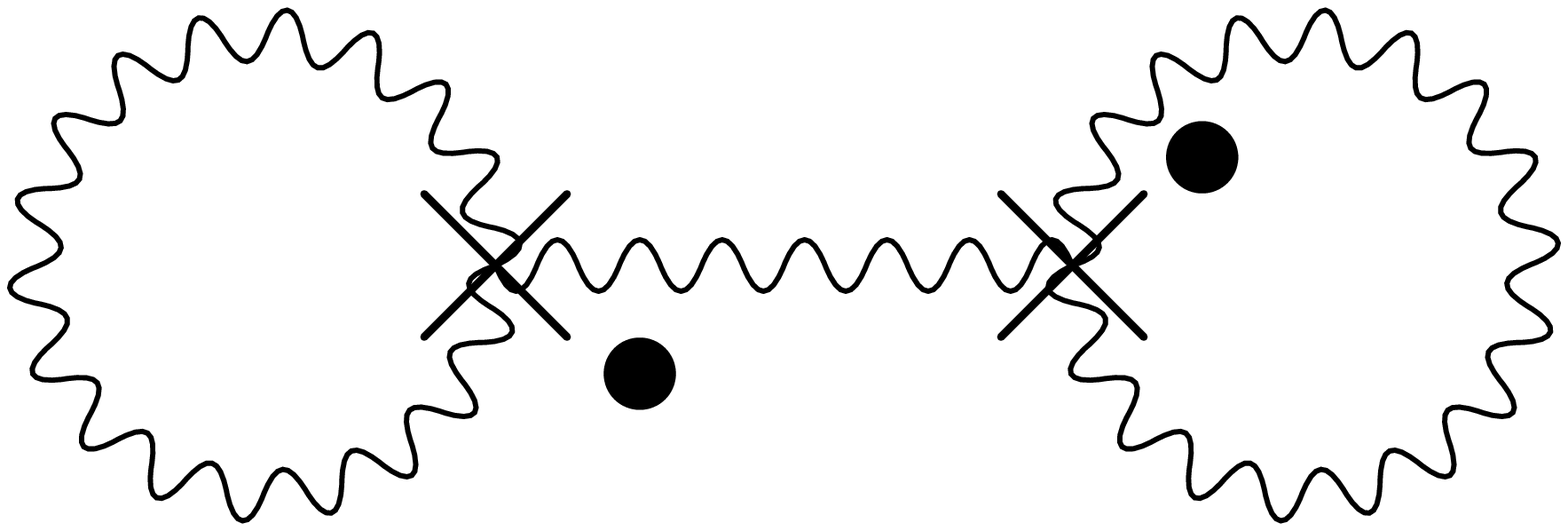,width=2cm}}
\ \ = \ \
\frac{1}{4}\ \ \raisebox{-.3cm}{\psfig{figure=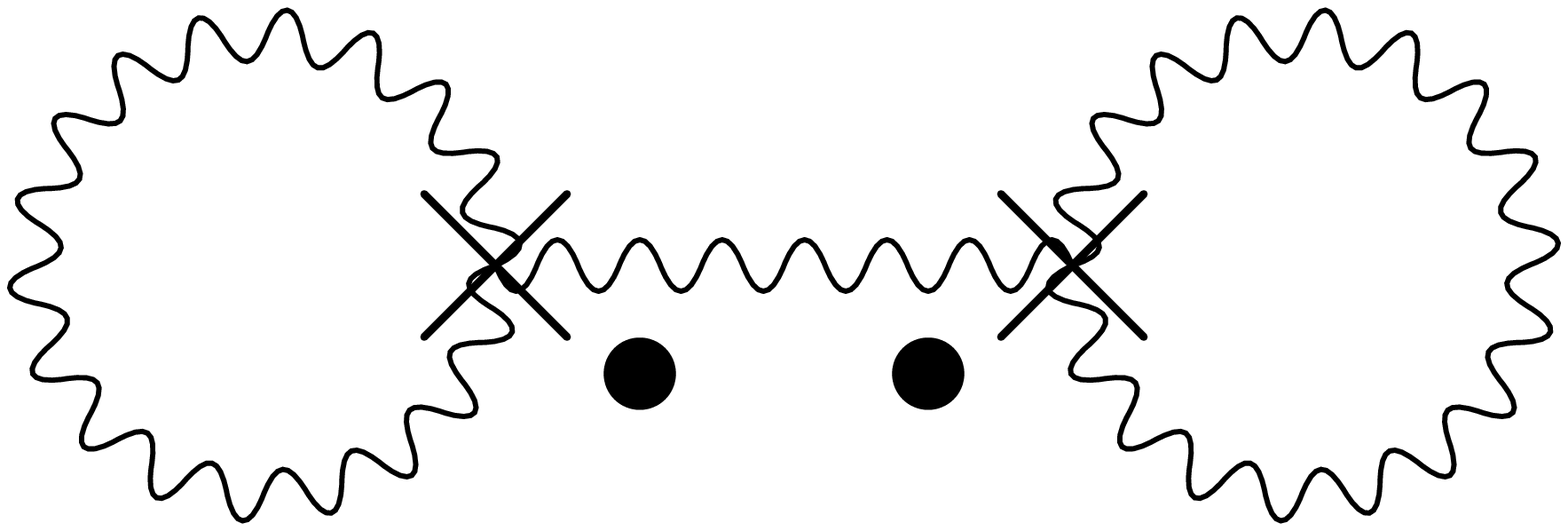,width=2cm}}
\ \ = \ \
\frac{1}{4}\ \ \raisebox{-1cm}{\psfig{figure=diagram11.ps,width=1cm}}
\ \ - \ \ \frac{1}{4}\
\left(\raisebox{-.5cm}{\psfig{figure=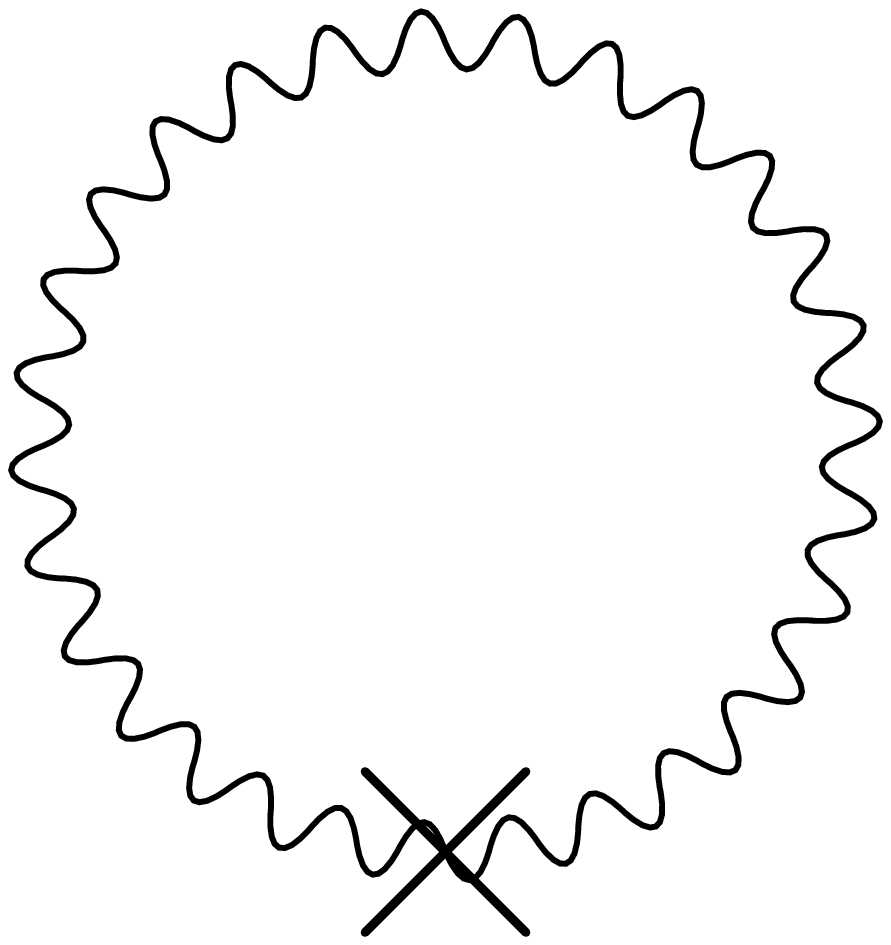,width=1cm}}\right)^{2}
\hspace{.7cm}
\\ \\ \hline \end{array}$$
Figure 1. Manipulation of graphs.\nonumber
\end{figure}

\begin{table}
$$
\begin{array}{|llll|}
\hline&&&\\
\raisebox{-.1cm}{\psfig{figure=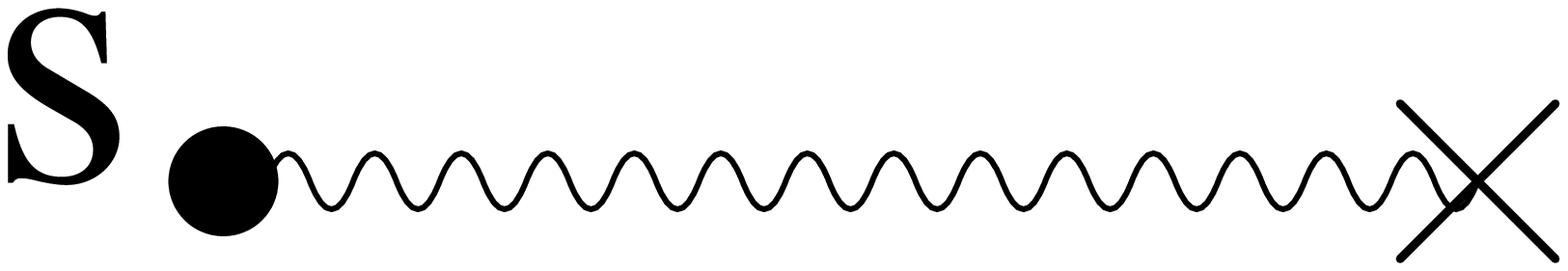,width=2cm}}=\frac{1}{2}s(1-s)&
\raisebox{0cm}{\psfig{figure=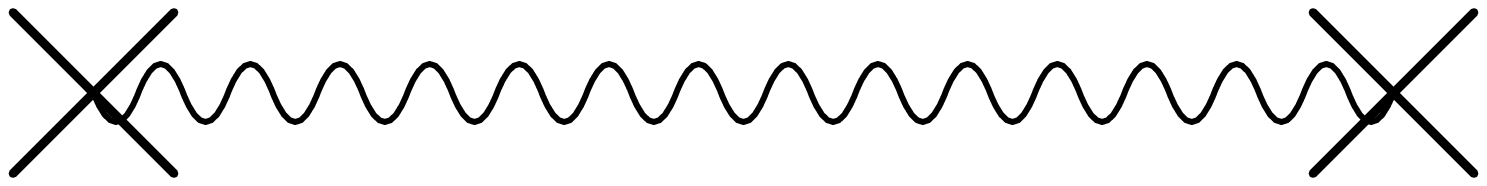,width=2cm}}=\frac{1}{12}&
\raisebox{-.5cm}{\psfig{figure=figure3.ps,width=1cm}}=\frac{1}{6}
\hspace{1cm}&
\raisebox{-.9cm}{\psfig{figure=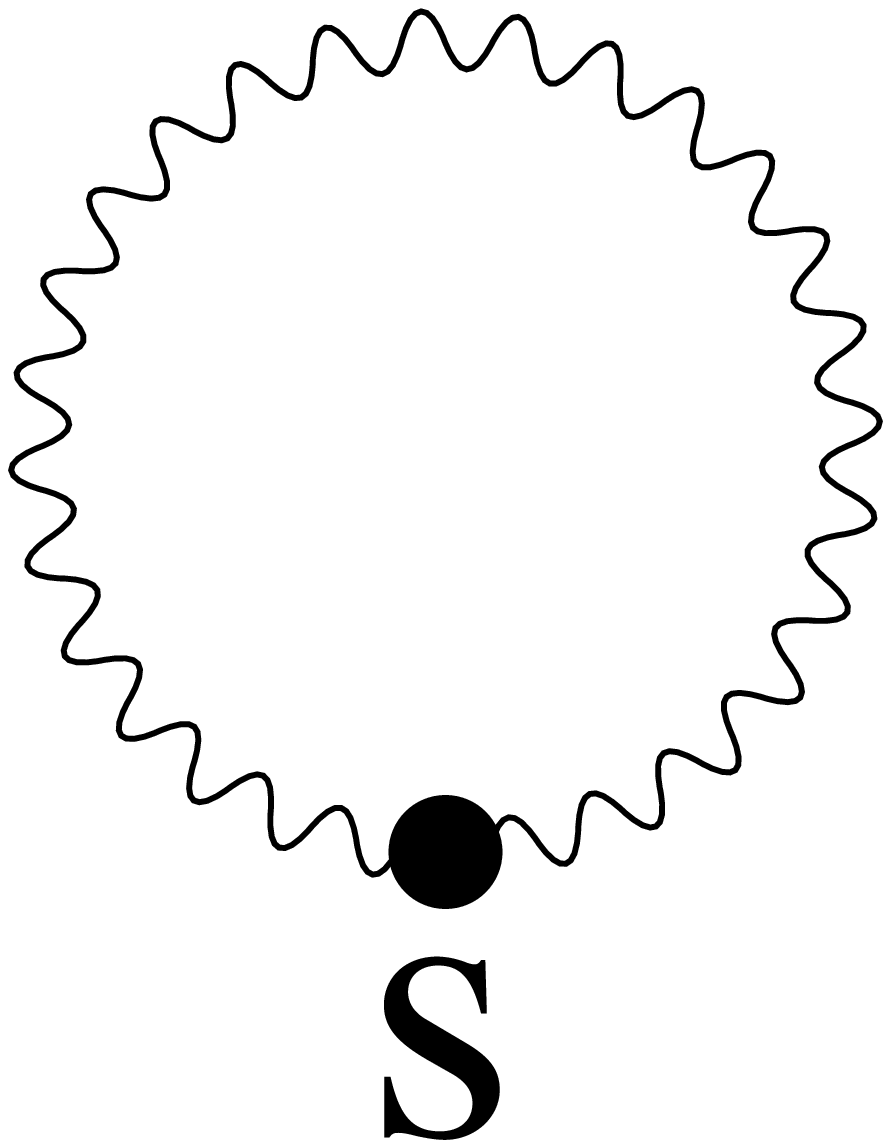,width=1cm}}=s(1-s)\\
\hline&&&\\
\raisebox{-.1cm}{\psfig{figure=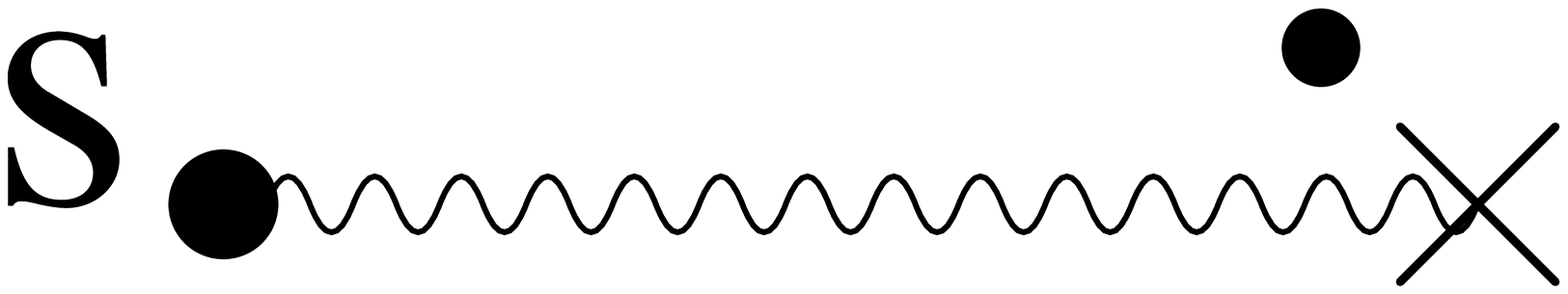,width=2cm}}=0&
\raisebox{-.15cm}{\psfig{figure=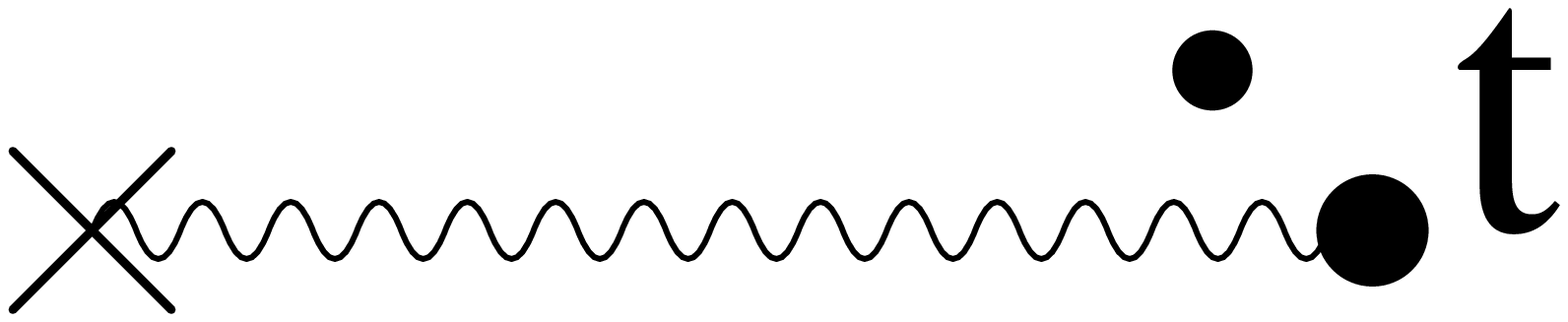,width=2.1cm}}=\frac{1}{2}-t&
\raisebox{-.5cm}{\psfig{figure=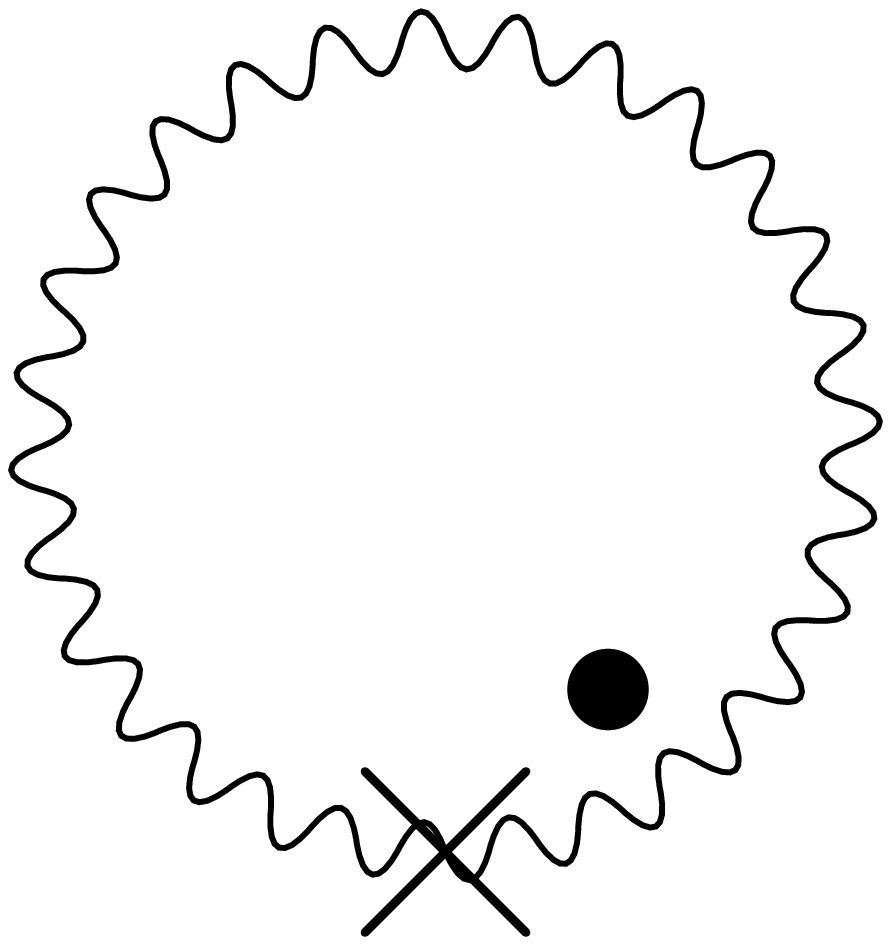,width=1cm}}=0&
\raisebox{-.9cm}{\psfig{figure=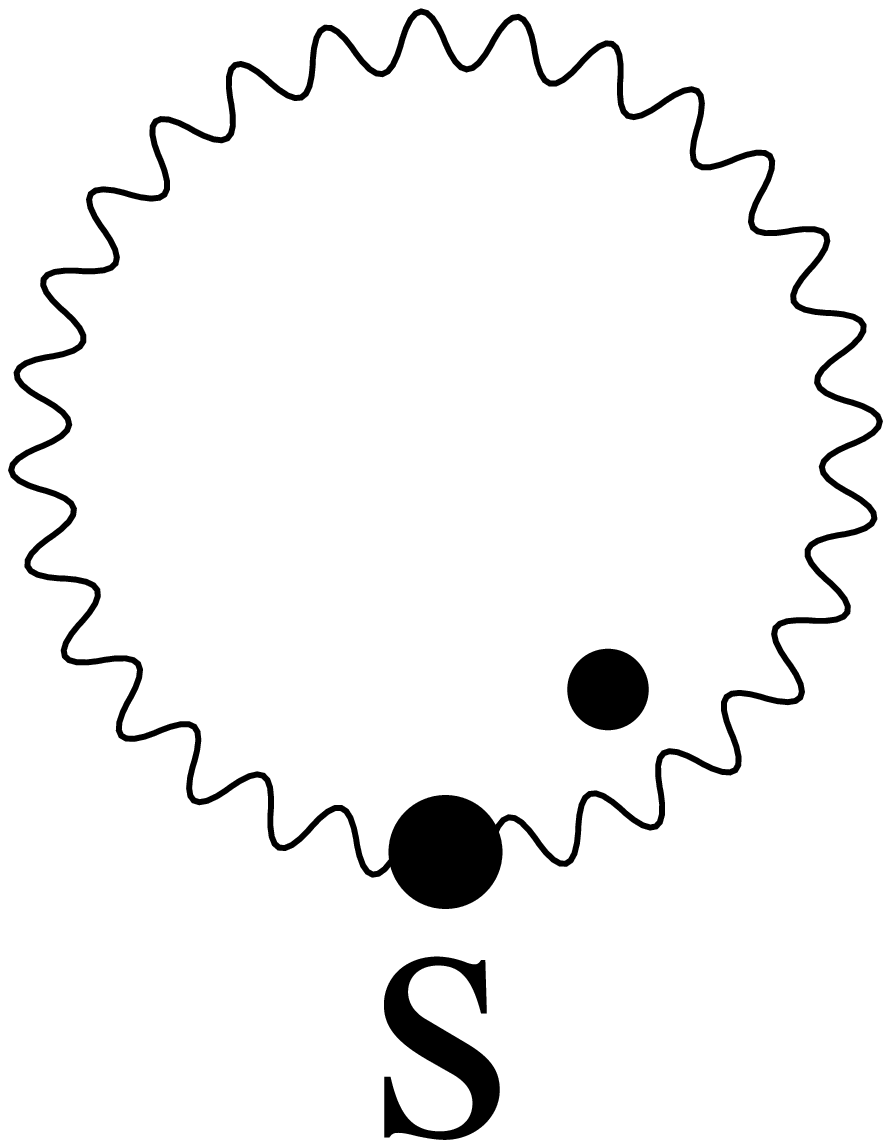,width=1cm}}=\frac{1}{2}-s
\\ \hline&&&\\
\multicolumn{4}{|l|}{\raisebox{-.1cm}{\psfig{figure=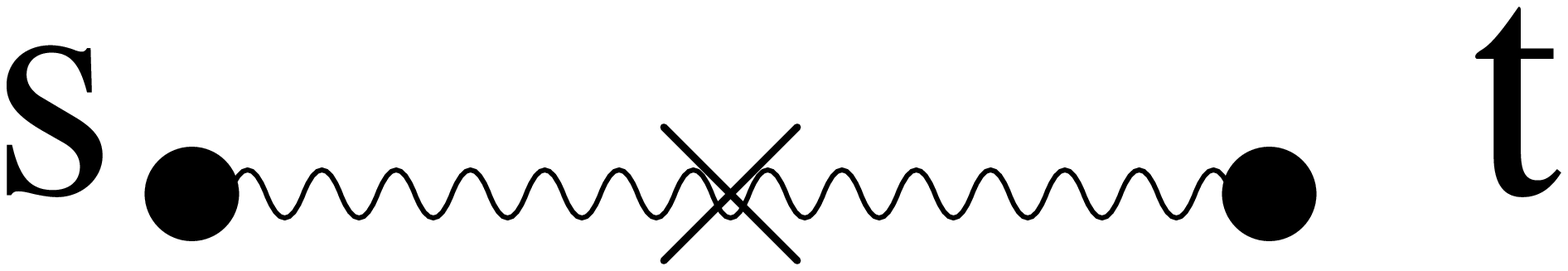,width=2.3cm}}\!=
\!\frac{1}{6}t(1-s)(2s-s^2-t^2)\theta(s-t)+
\frac{1}{6}s(1-t)(2t-t^2-s^2)\theta(t-s)}\\
\raisebox{-.1cm}{\psfig{figure=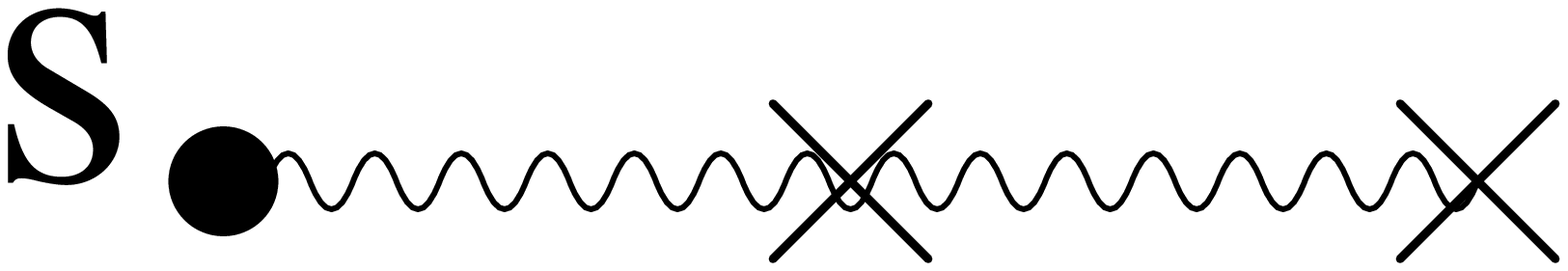,width=2cm}}
=\!\frac{1}{24}s(1\!-\!s)(1\!+\!s\!-\!s^2)&
\raisebox{0cm}{\psfig{figure=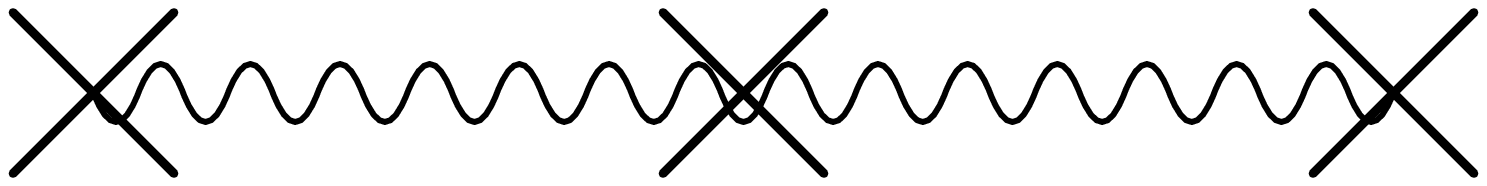,width=2cm}}=\frac{1}{120}&
\raisebox{-.5cm}{\psfig{figure=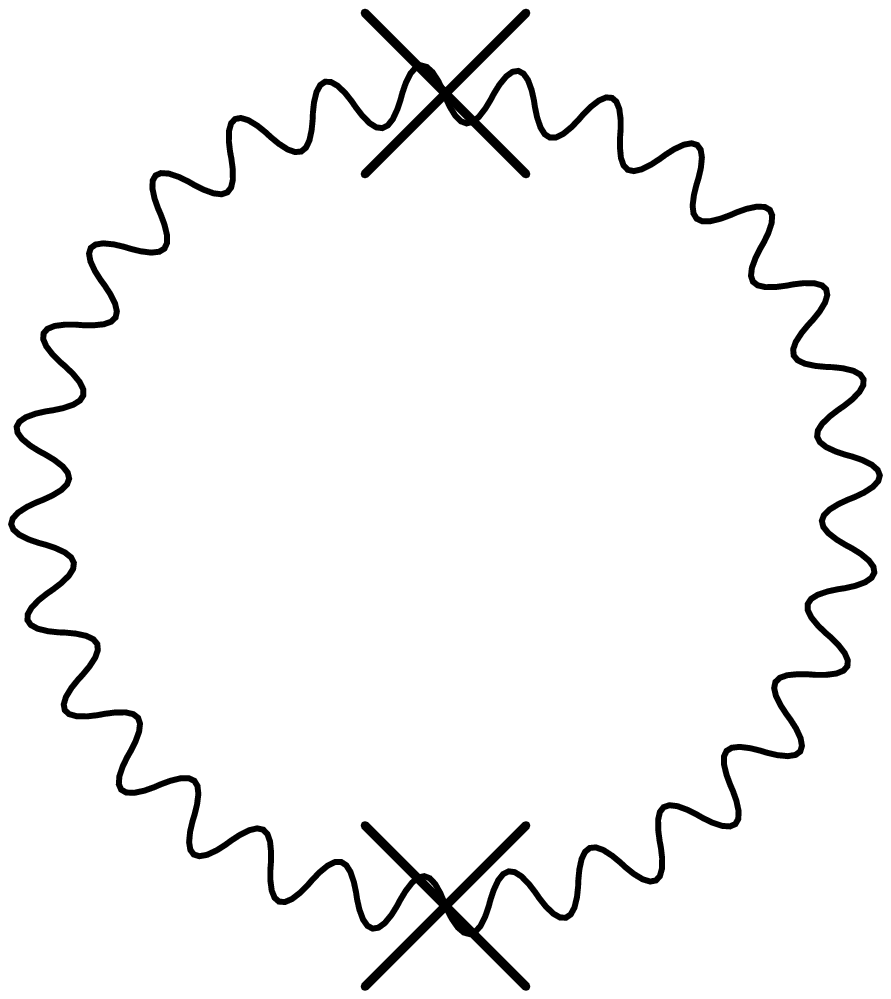,width=1cm}}=\frac{1}{90}&
\raisebox{-.9cm}{\psfig{figure=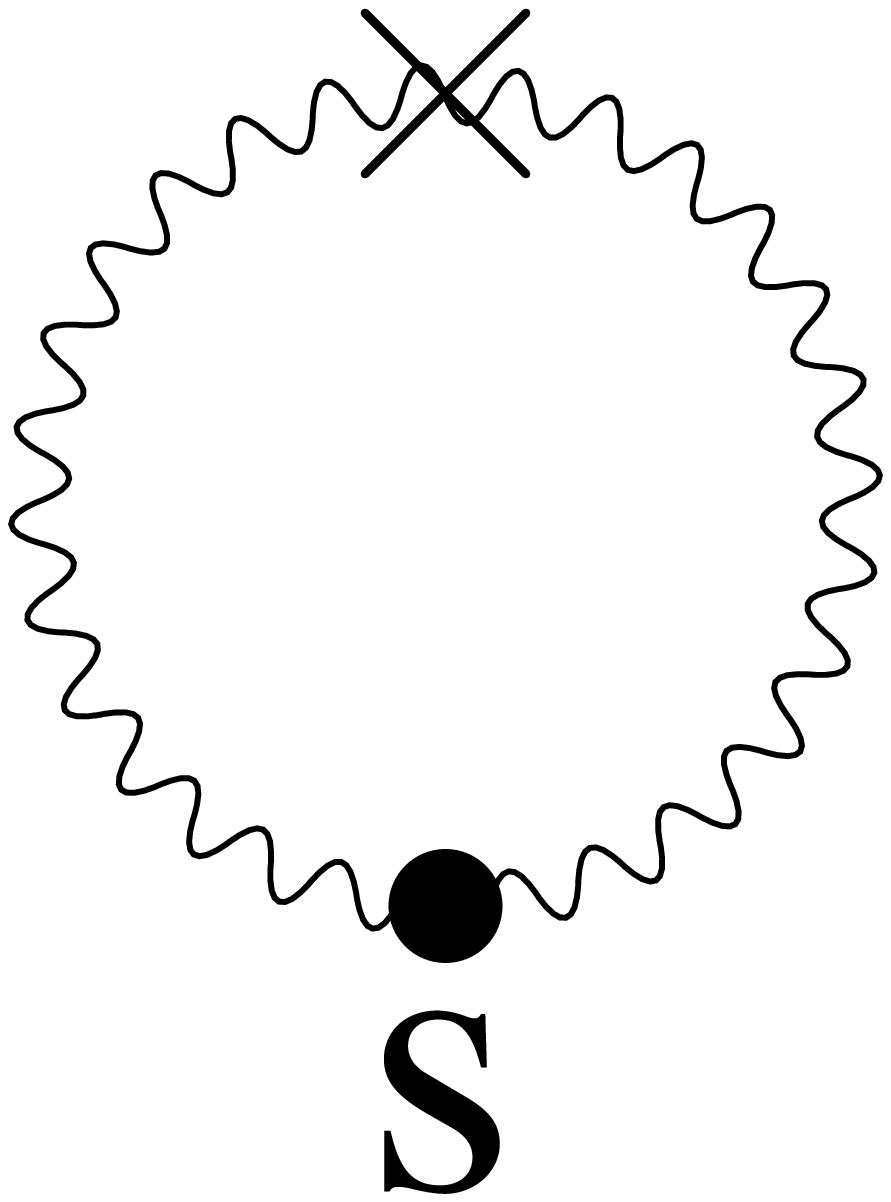,width=1cm}}=
\frac{1}{3}s^2(1\!-\!s)^2\\
\hline&&&\\
\multicolumn{4}{|l|}{\raisebox{-.1cm}{\psfig{figure=diagram8.ps,width=2.3cm}}
\!=\!\frac{1}{2}(s\!-\!t)(t(1\!-\!s)\theta(s\!-\!t)\!+\!
s(1\!-\!t)\theta(t-s))\!=\!\frac{1}{2}(s\!-\!t)\Delta(s,t)
\!=\!
-\left(\raisebox{-.1cm}{\psfig{figure=diagram9.ps,width=2.3cm}}\! \right)}\\
\raisebox{-.1cm}{\psfig{figure=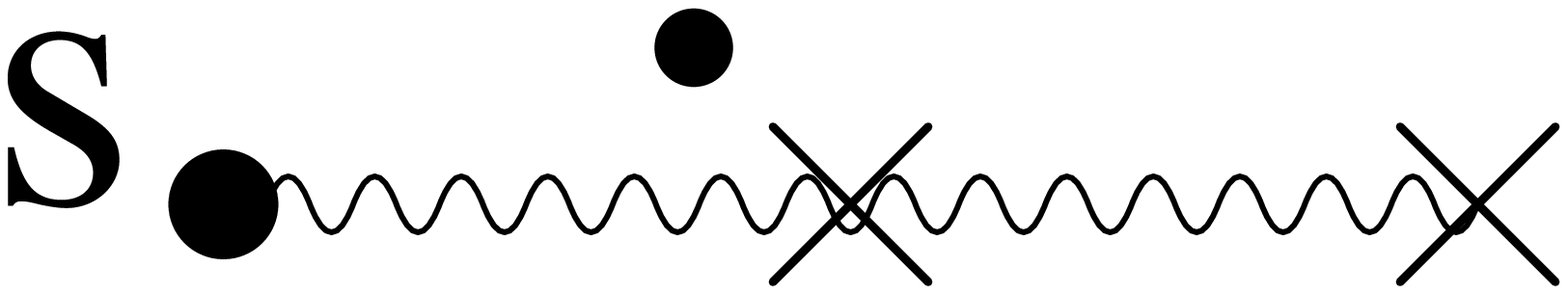,width=2cm}}
=\frac{1}{12}s(1-s)(2s-1)&
\raisebox{0cm}{\psfig{figure=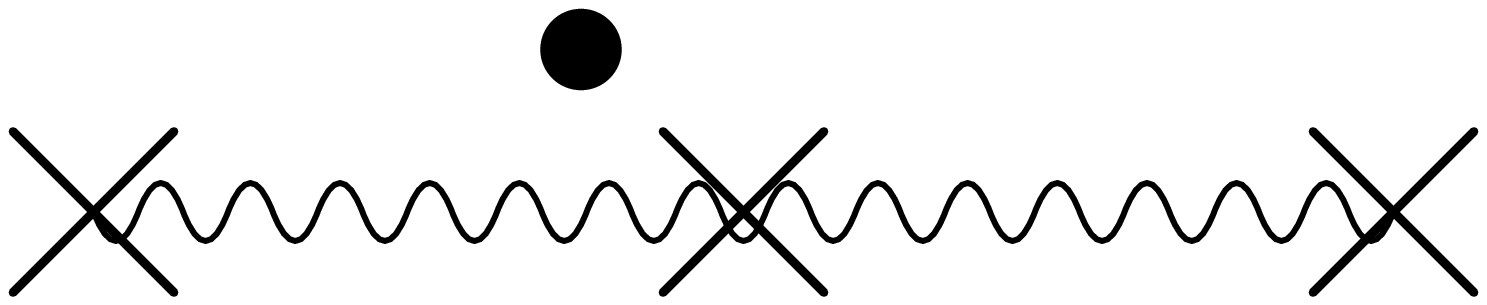,width=2cm}}=0&
\raisebox{-.5cm}{\psfig{figure=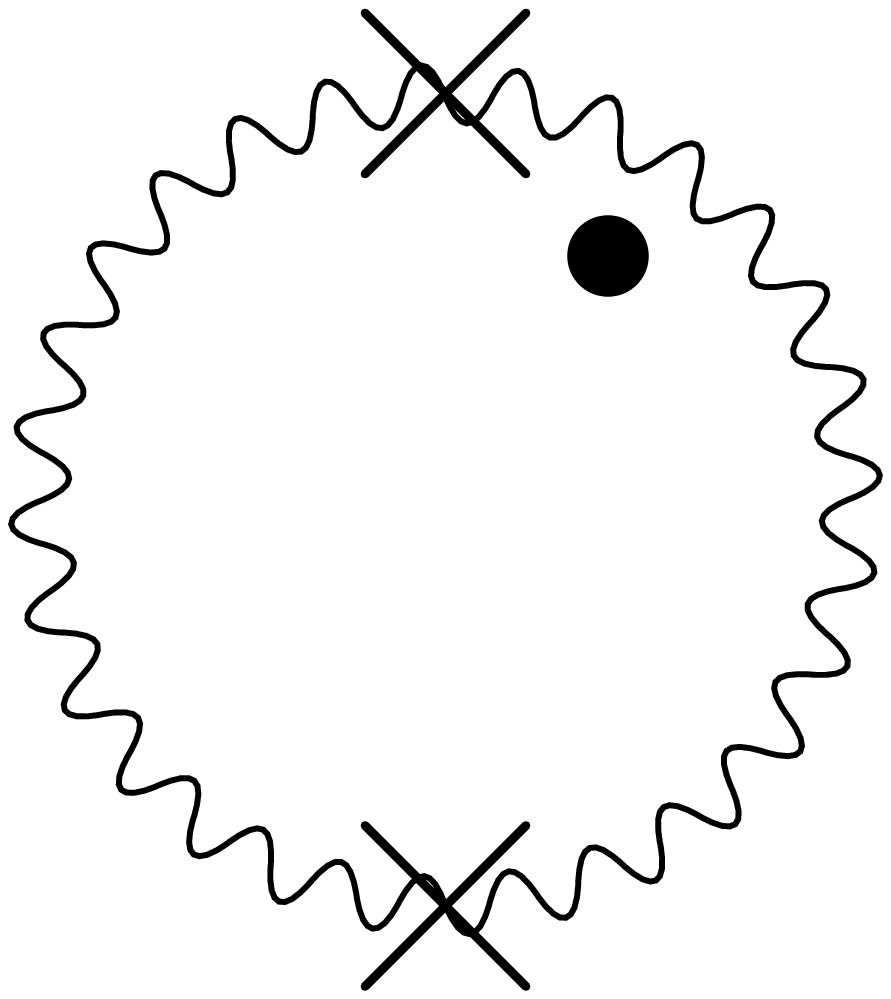,width=1cm}}=0&
\raisebox{-.9cm}{\psfig{figure=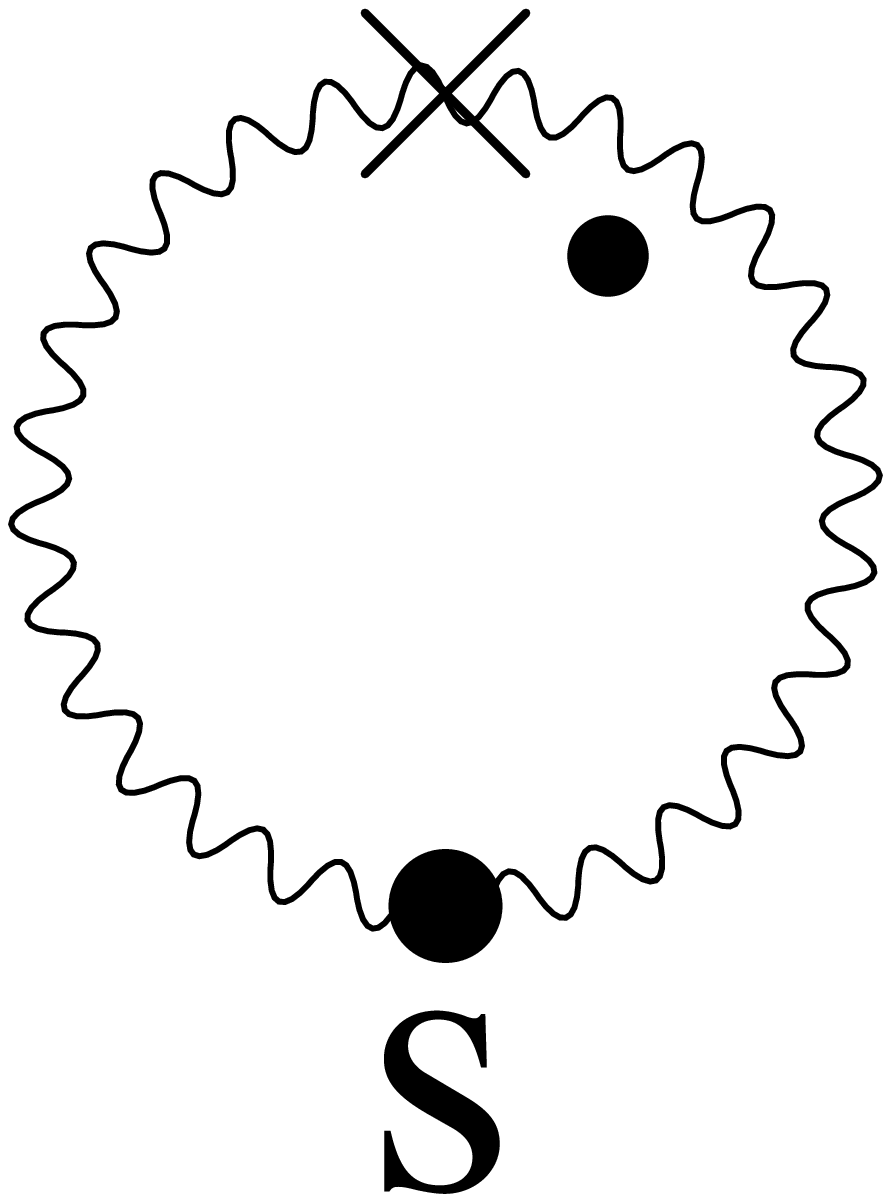,width=1cm}}=0\\
\hline&&&\\
\multicolumn{4}{|l|}{\raisebox{-.1cm}{\psfig{figure=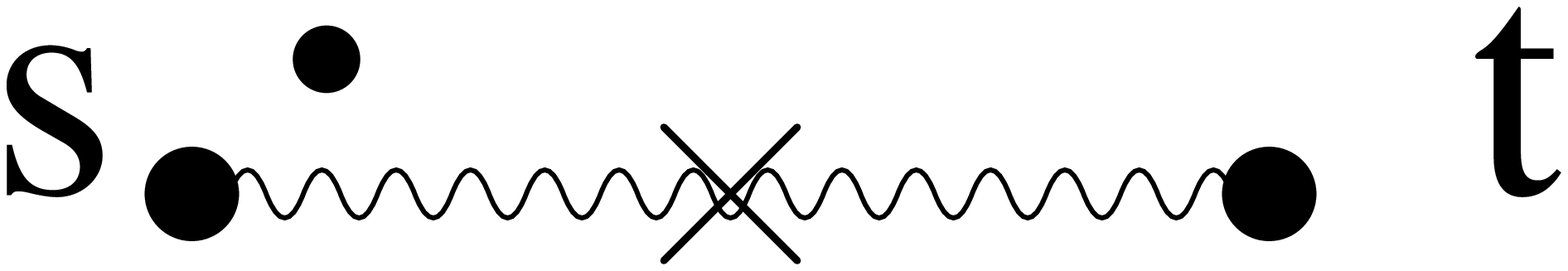,width=2.3cm}}=
\frac{1}{6}t(t^2+3s^2+2-6s)\theta (s-t)+
\frac{1}{6}(1-t)(2t-t^2-3s^2)\theta (t-s)}\\
\multicolumn{2}{|l}{\raisebox{-.1cm}{\psfig{figure=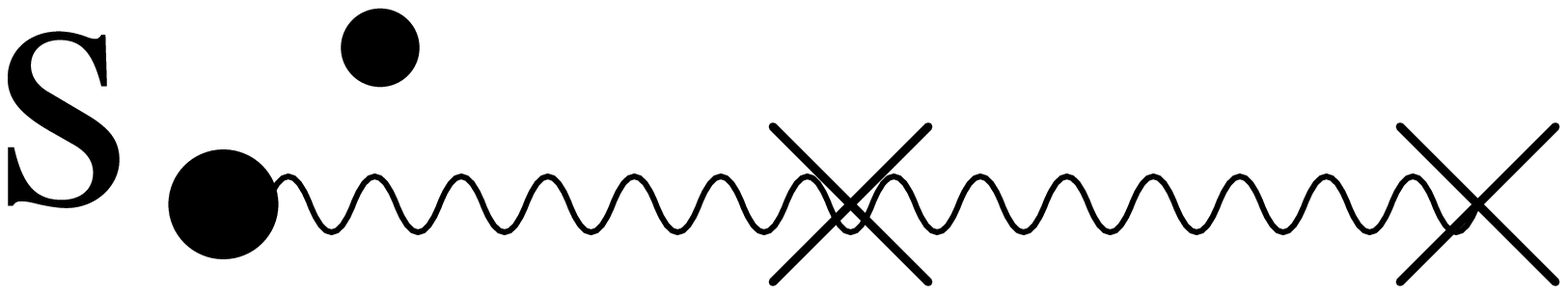,width=2cm}}
=\frac{1}{24}(2s-1)(2s^2-2s-1)}&
\multicolumn{2}{l|}{\raisebox{-.9cm}{\psfig{figure=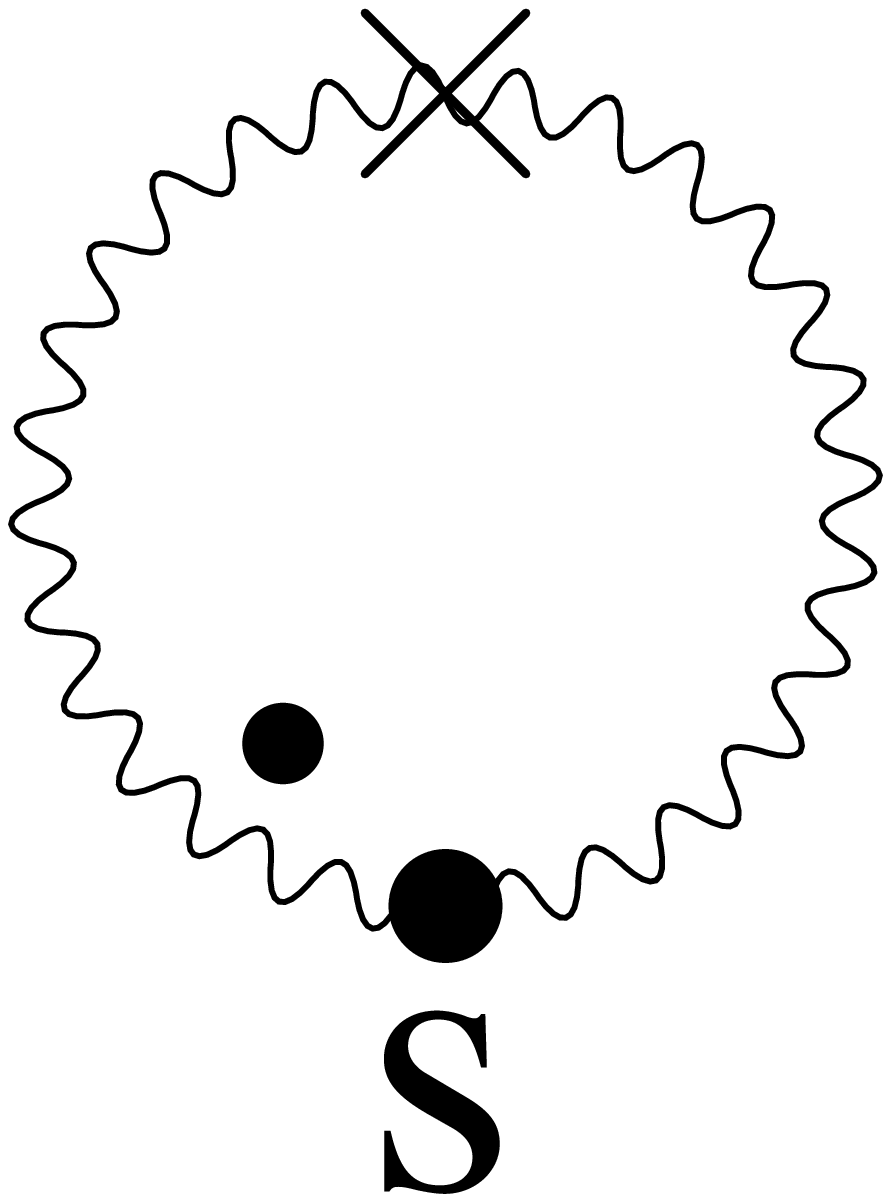,width=1cm}}
=\frac{1}{3}s(1-s)(1-2s)}\\
\hline&&&\\
\ \ \raisebox{-.5cm}{\psfig{figure=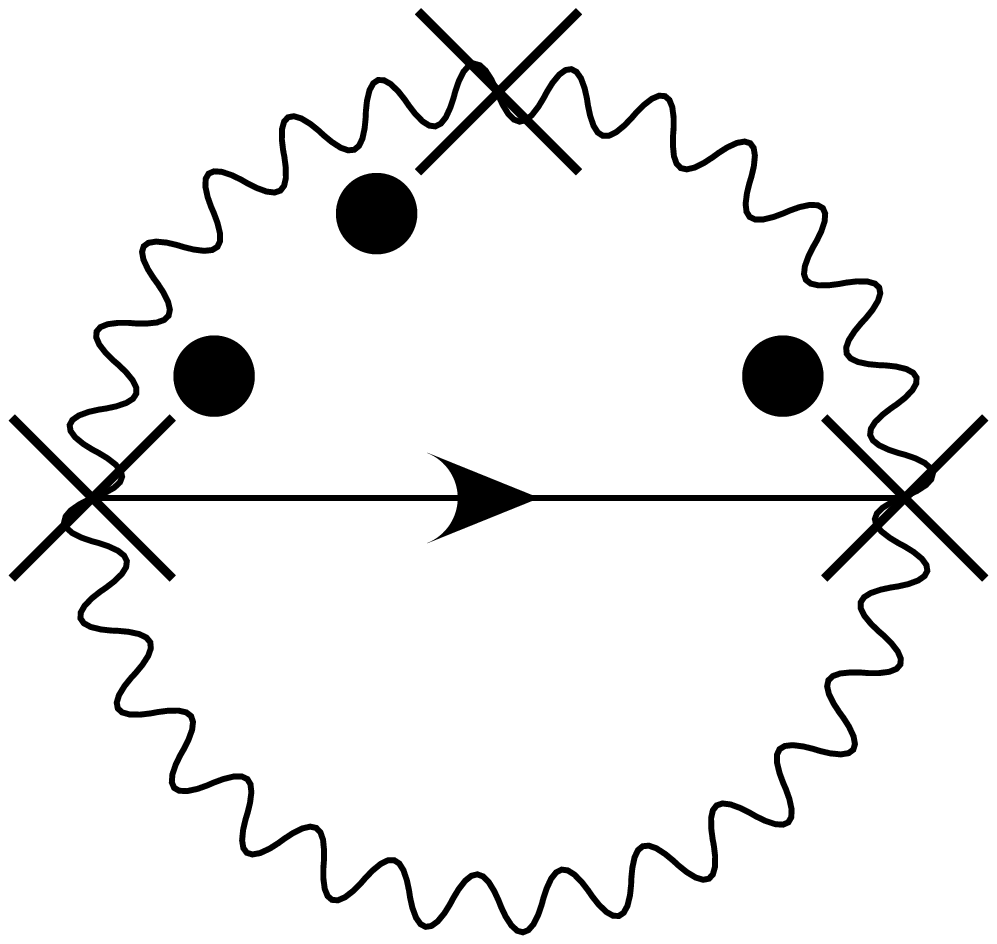,width=1.2cm}}=\frac{1}{80}&
\raisebox{-.5cm}{\psfig{figure=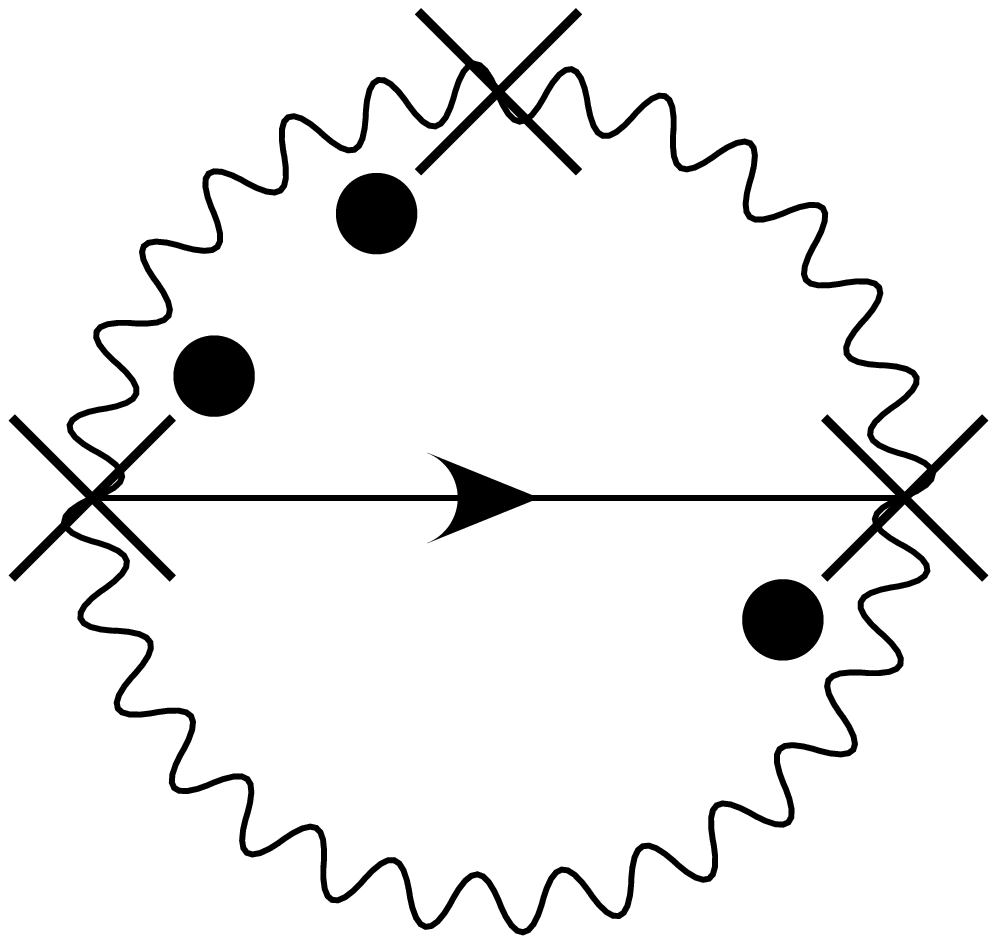,width=1.2cm}}=\frac{1}{240}&
\raisebox{-.5cm}{\psfig{figure=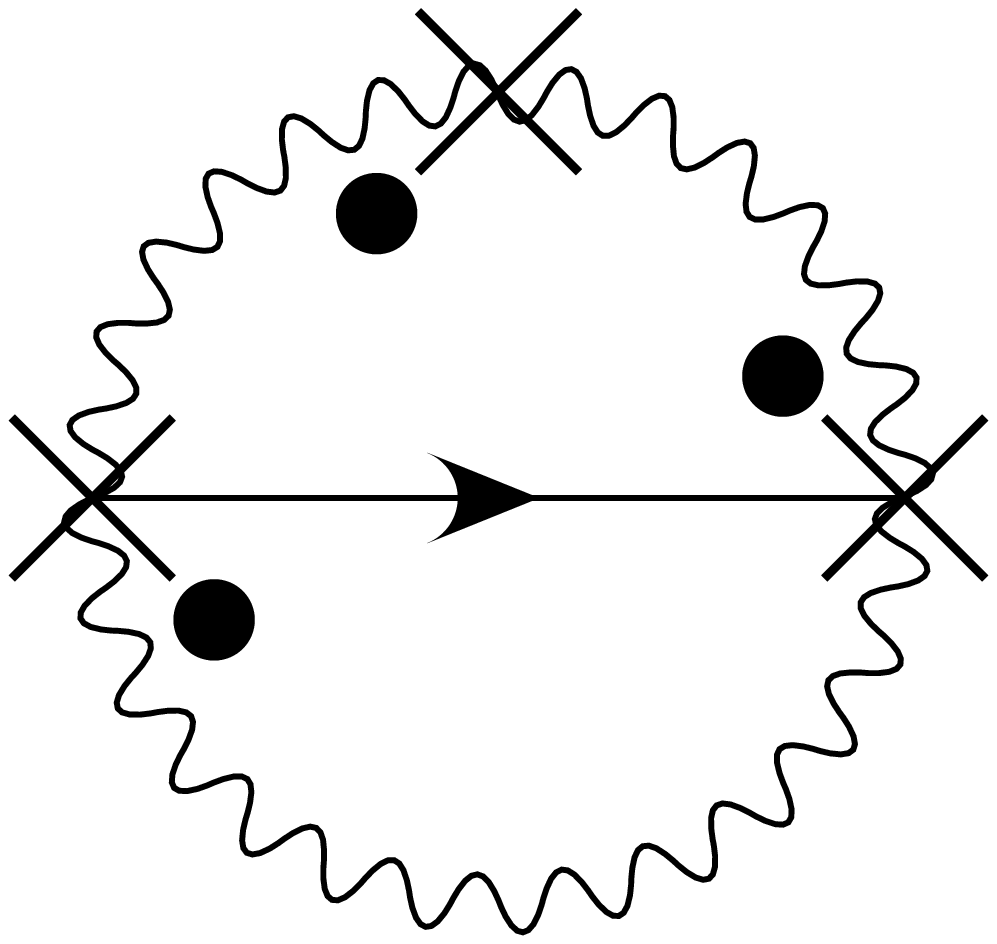,width=1.2cm}}=\frac{1}{240}&
\raisebox{-.5cm}{\psfig{figure=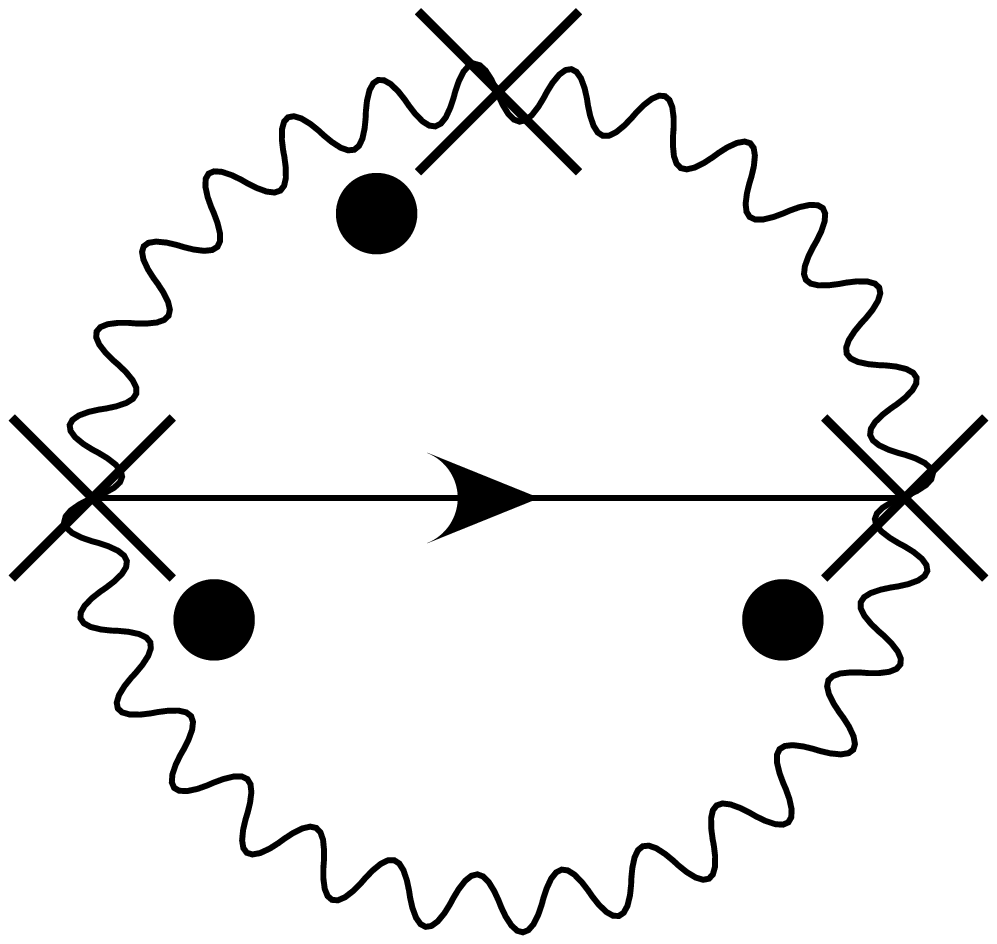,width=1.2cm}}=-\frac{1}{240}\\
&&&\\ \hline
\end{array}
\nonumber $$
\nonumber
Table 1. ``Table of Integrals''.\nonumber
\end{table}

Before considering the graphs required to calculate the anomaly ${\cal A}$,
let us discuss one more property of the
propagators~(\ref{prop1})-(\ref{prop3}). Suppose we preferred the convention
of integrating $\int_{-1}^{0}$ to $\int_{0}^{1}$, then we could convert
our results via the variable change $s^\prime=-s$ or $s^\prime=s-1$
under both of which $\int_{0}^{1}ds\rightarrow \int_{-1}^{0}ds$.
However, although the propagators $\Delta(s,t)$ and
$^\cdot\!\Delta\!^\cdot(s,t)$ transform identically  under both of
these relabellings,
the  $q\dot{q}$ propagator transforms with a relative sign
between the two
variable changes.
Hence in any graph ${\cal G}=\int_{0}^{1}(\prod_{i}ds_i)(\prod \Delta)
(\prod^{N}\Delta\!^\cdot)(\prod \ ^\cdot\!\! \Delta\!^\cdot)$, by subsequently
changing variables $s^\prime=s-1$ and  $s^{\prime\prime}=-s^\prime$ we
have ${\cal G}=(-)^N{\cal G}$ so that only graphs with an even number of
$q\dot{q}$ propagators are non-vanishing.
Indeed study of possible bosonic graphs (graphs without fermion propagators)
shows that the vertex $\frac{-1}{2}\partial^n_{\alpha_1\cdots\alpha_n}
\omega_{\mu ab}\Psi^a\Psi^b\frac{1}{n!}\int_{0}^{1}
q^{\alpha_1}\cdots q^{\alpha_n} \dot{q}^\mu$ must appear an even number
of times\footnote{This is shows that bosonic graphs must have
$4k+4,\ \ k=0,1,\ldots \ $ external $\Psi$'s which is consistent with the
chiral anomaly existing in $4k+4$ dimensions only.} $\ \ $ so that any
bosonic graph has an even number of ``dots'' and therefore
an even number of $q\dot{q}$ propagators.
Furthermore notice that $\epsilon(s-t)$ changes sign under $s^\prime= -s$
but is invariant under $s^\prime=s-1$ so that we can now argue that
the total number of $q\dot{q}$ propagators plus the number of
$\epsilon(s-t)$'s in
any graph must be even. Therefore any non-vanishing graph containing
$\epsilon(s-t)$ vanishes when a single $\epsilon(s-t)$ is replaced by $K^{ab}$.
In fact one finds via this argument (or directly) that all two loop graphs
involving $K^{ab}$ vanish separately. At higher loops only an even number
of $K^{ab}$'s can appear (since the total number of
$q\dot{q}$ and fermion propagators must be even
for any graph with $4k+4$ external
$\Psi$'s) and must cancel amongst themselves or as
contractions on symmetric invariants.

Using the results in table 1 and the vertices and propagators given above
one may now write down all relevant graphs and their results by inspection
(at worst one may need to perform a single integral $\int_{0}^{1}$ after
appropriate integrations by parts and manipulations as explained above, see
also figure 1 for examples).
We work in $n=4$ dimensions so all graphs must have four external fermions
to saturate the Grassmann integral $\int d^4\Psi$. To calculate
$Z[q^\mu,\Psi^a]$ we must include all graphs, disconnected, connected and
one-particle irreducible. At one loop order we find
\be
\raisebox{-.5cm}{\psfig{figure=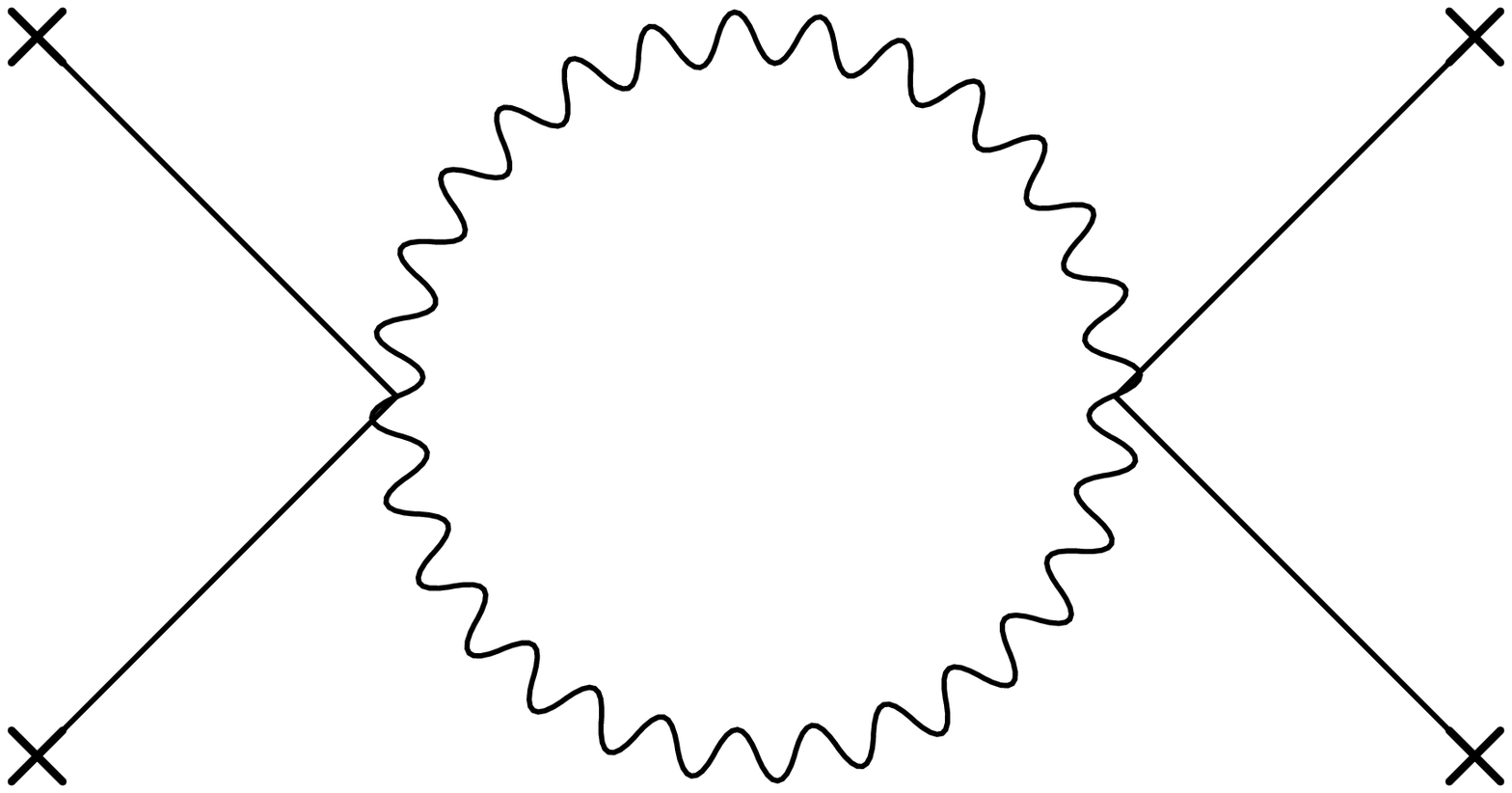,width=2cm}}\ \
=\frac{1}{192}R_{\mu\nu ab}R^{\mu\nu}\0_{cd}\Psi^a\Psi^b\Psi^c\Psi^d.
\ee
For brevity we leave the dots off the graphs, to reinstate them one
must write all independent combinations of dots allowed by the
vertices~(\ref{vertices}). Here there are two such independent graphs,
however using the integration by parts given in\r{elvis} and
anti-symmetry of $R_{\mu\nu ab}$ in $\mu$ and $\nu$ one needs only
calculate two times one of them. With this result we can already calculate
the gravitational chiral anomaly using\r{anom} and find the correct
result
\be
{\cal A}=-\frac{1}{384\pi^2}\int
d^4y\sqrt{g}R_{\mu\nu ab}\left( \frac{1}{2}\epsilon^{abcd}
R^{\mu\nu}\0_{cd}\right) .
\ee
At two loops the results are
\bea
\left(\raisebox{-.5cm}{\psfig{figure=diagram15.ps,width=2cm}}
\times\raisebox{-1cm}{\psfig{figure=diagram11.ps,height=2cm}}\right)
&=&\frac{-1}{4608}RR_{\mu\nu ab}R^{\mu\nu}\! \ _{cd}\Psi^a\Psi^b\Psi^c\Psi^d\\
\nonumber \\
\raisebox{-.5cm}{\psfig{figure=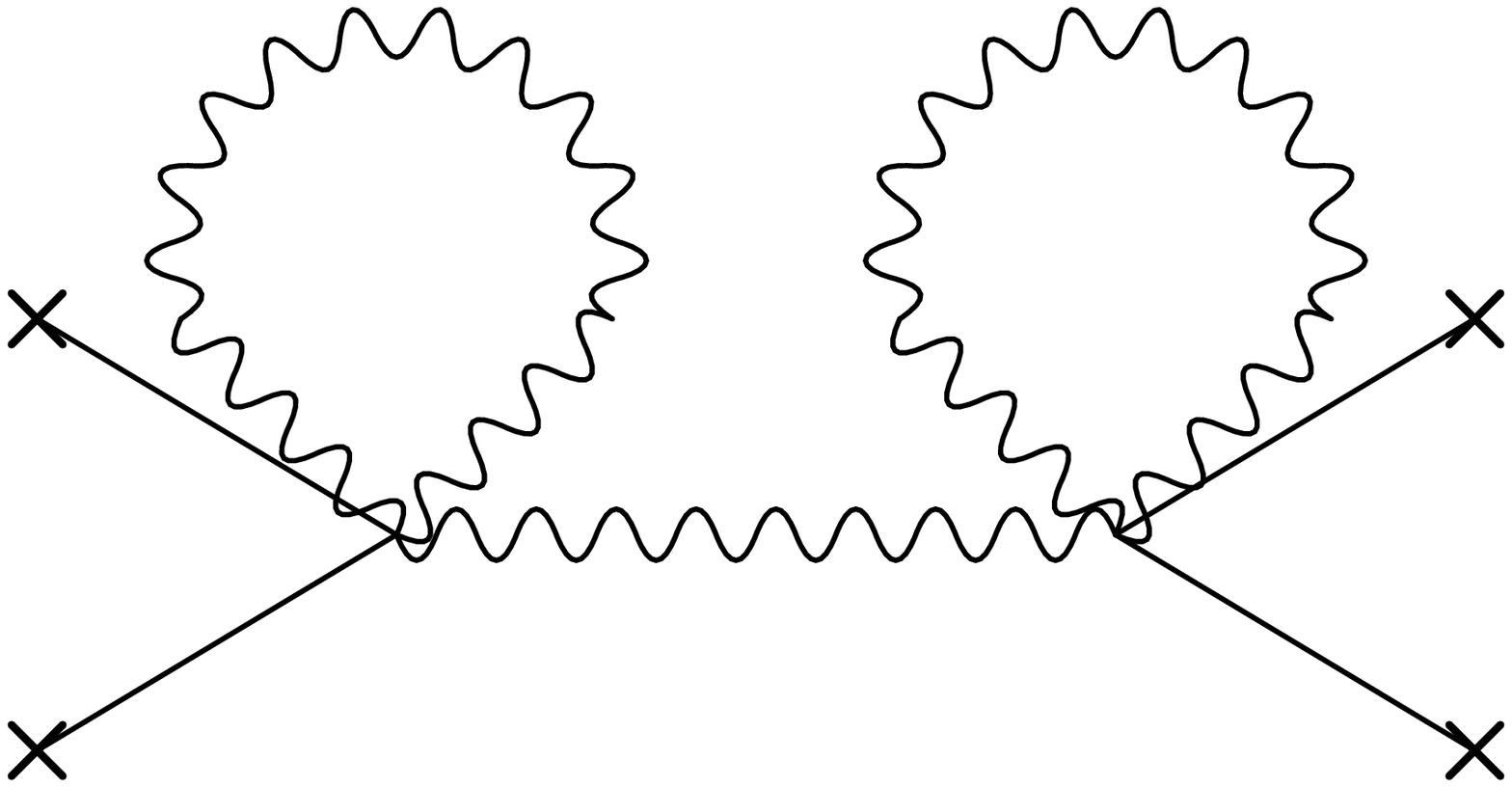,width=2cm}}\hspace{.4cm}
&=&\frac{1}{5760}D_\alpha R^{\alpha\mu}\0 _{ab}D_\beta R^\beta\0_{\mu cd}
\Psi^a\Psi^b\Psi^c\Psi^d\\ \nonumber \\
\raisebox{-.5cm}{\psfig{figure=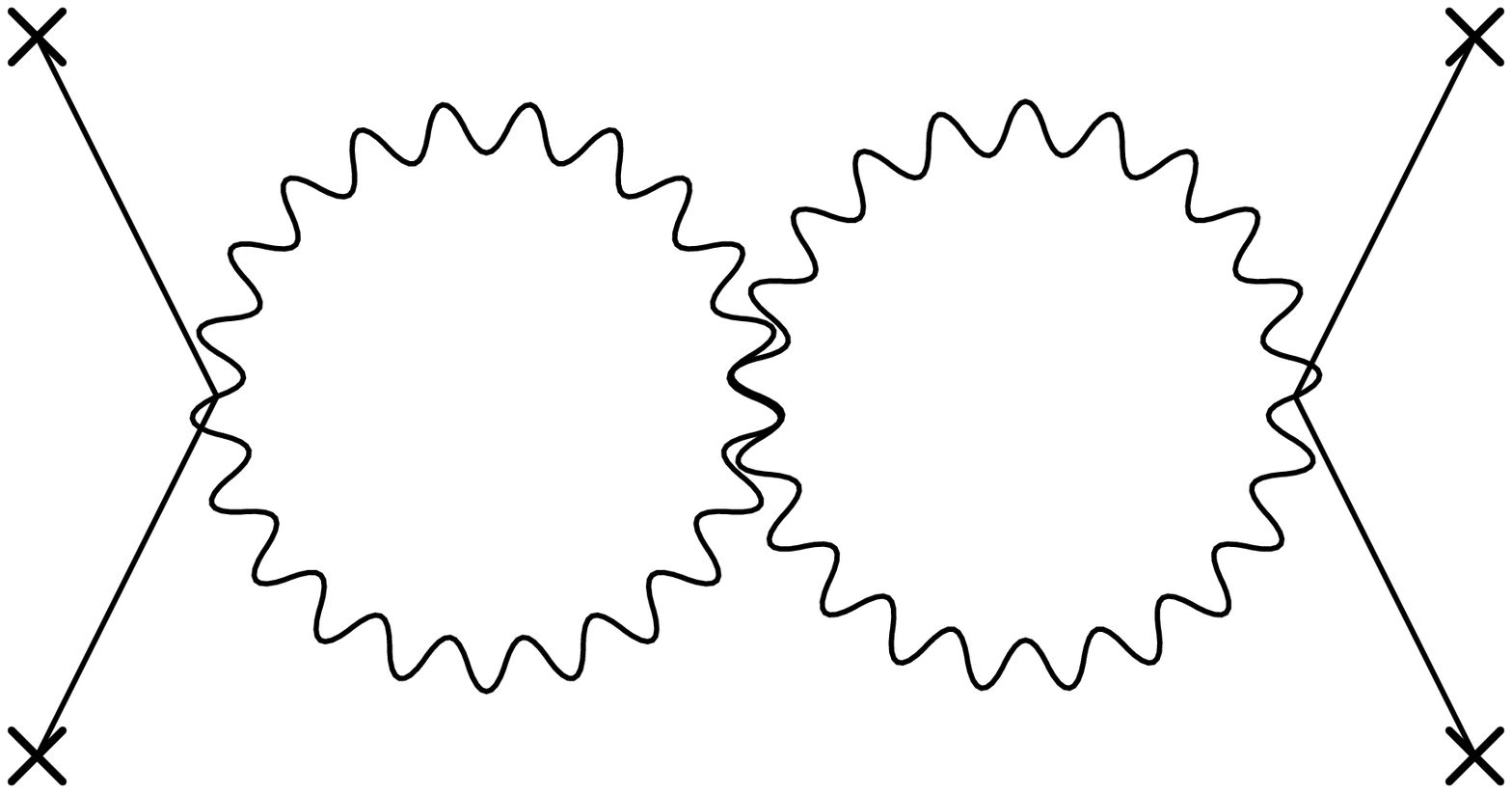,width=2cm}}\hspace{.4cm}
&=&-\frac{1}{2880}R^{\mu(\nu\sigma)\rho}R_{\mu\nu ab}R_{\rho\sigma cd}
\Psi^a\Psi^b\Psi^c\Psi^d\\ \nonumber \\
\raisebox{-.5cm}{\psfig{figure=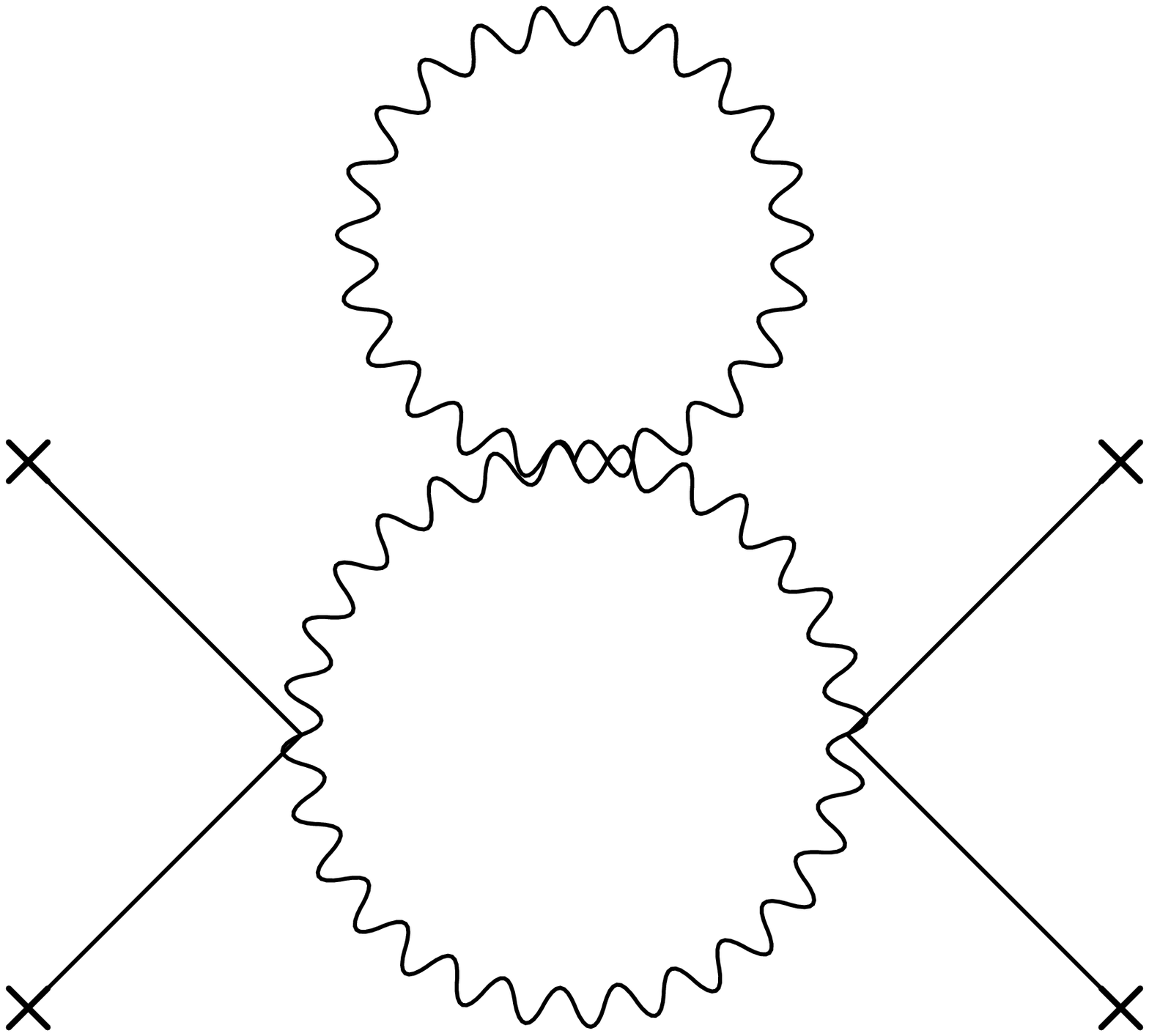,width=2cm}}\hspace{.4cm}
&=&\frac{1}{2880}R_{\mu\nu}R^{\mu\alpha}\0_{ab}R^\nu\0_{\alpha cd}
\Psi^a\Psi^b\Psi^c\Psi^d\\ \nonumber \\
\raisebox{-.5cm}{\psfig{figure=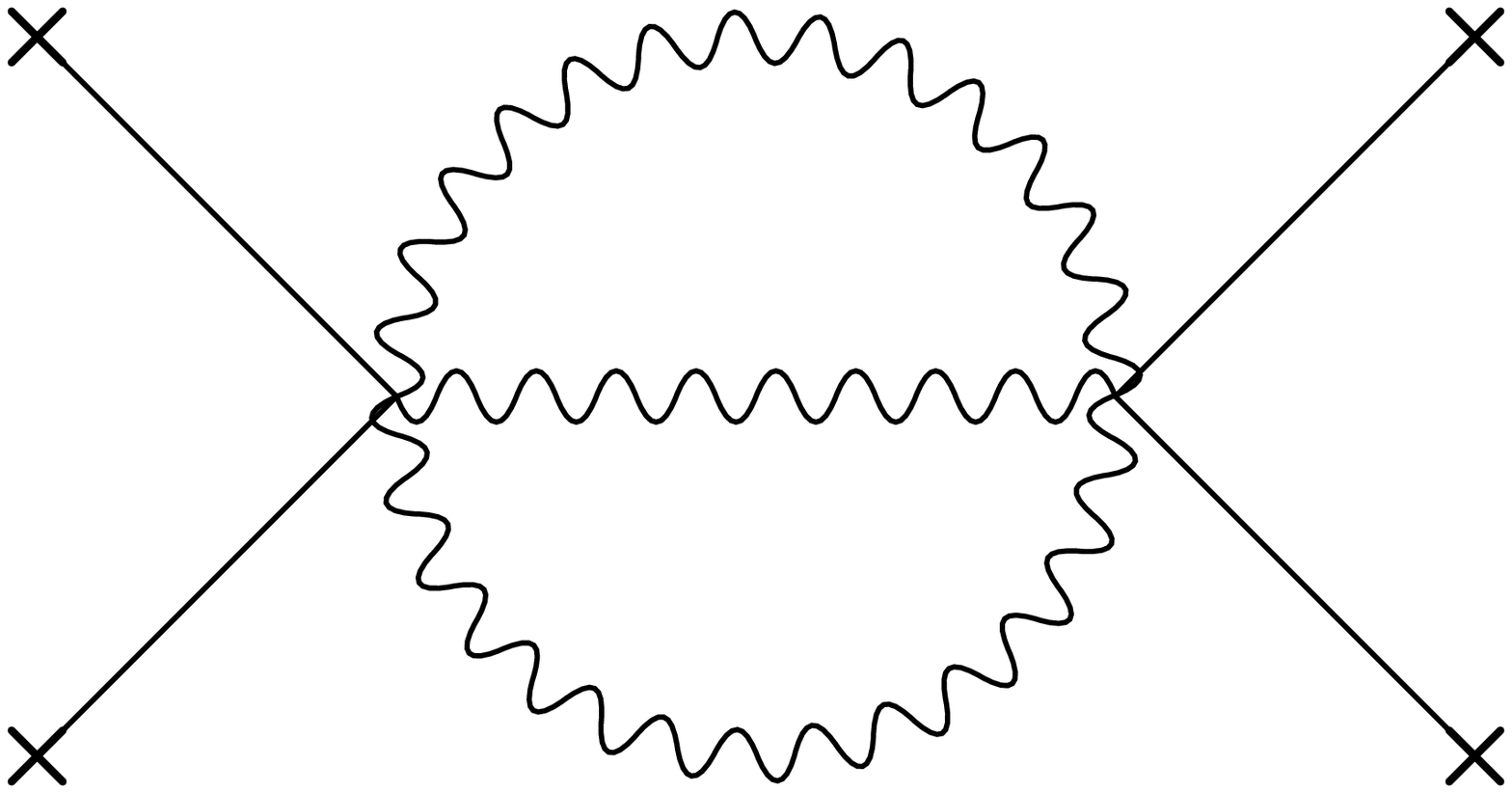,width=2cm}}\hspace{.4cm}
&=&\frac{1}{1440}D_\alpha R_{\mu\nu ab}D^\alpha R^{\mu\nu}\0_{cd}
\Psi^a\Psi^b\Psi^c\Psi^d\\ \nonumber \\
\raisebox{-.5cm}{\psfig{figure=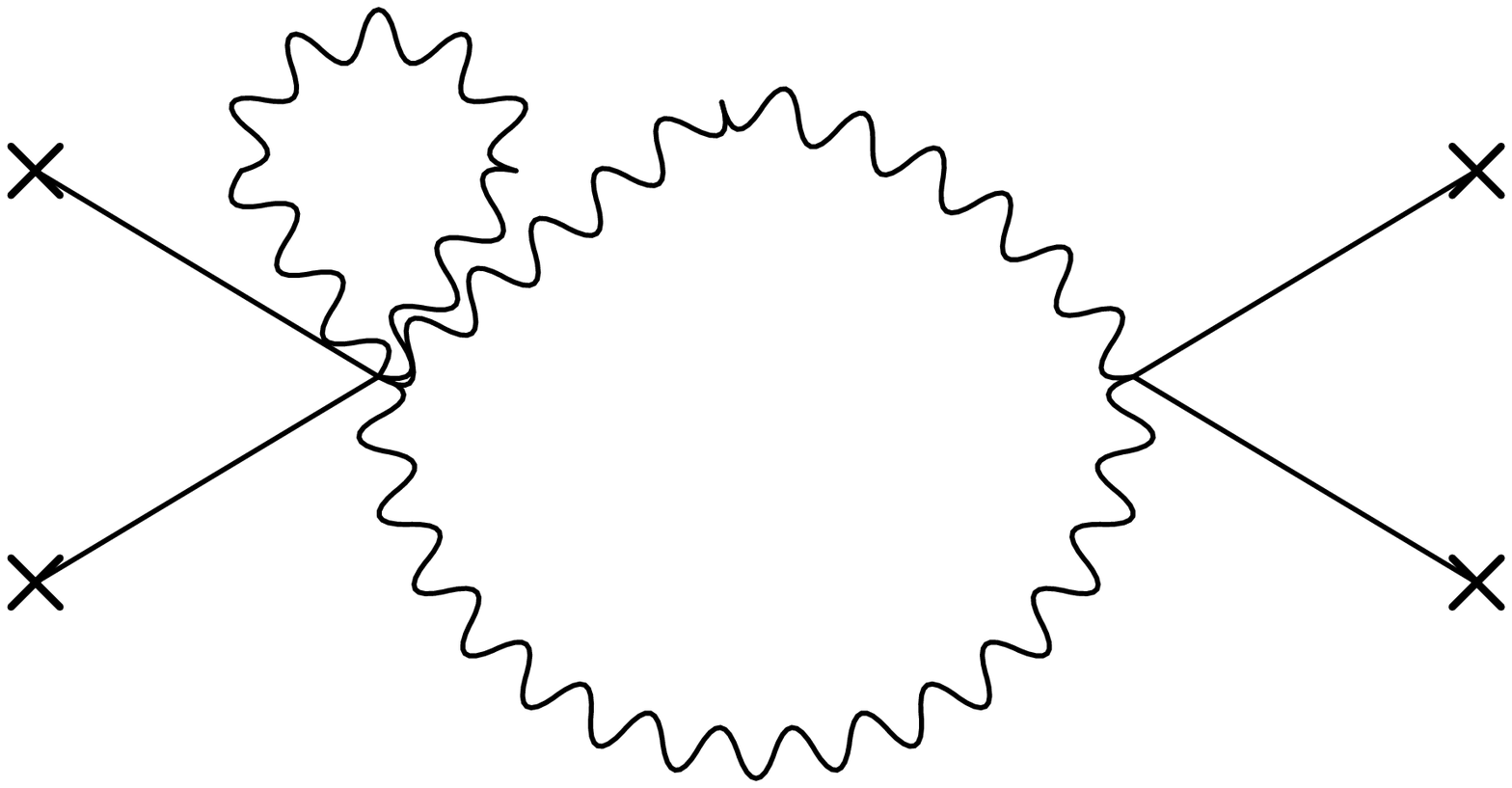,width=2cm}}\hspace{.4cm}
&=&\frac{1}{960}\Box\partial_{[\mu}\omega_{\nu]ab}R^{\mu\nu}\0_{cd}
\Psi^a\Psi^b\Psi^c\Psi^d\label{noncovariant}\\ \nonumber \\
\raisebox{-.5cm}{\psfig{figure=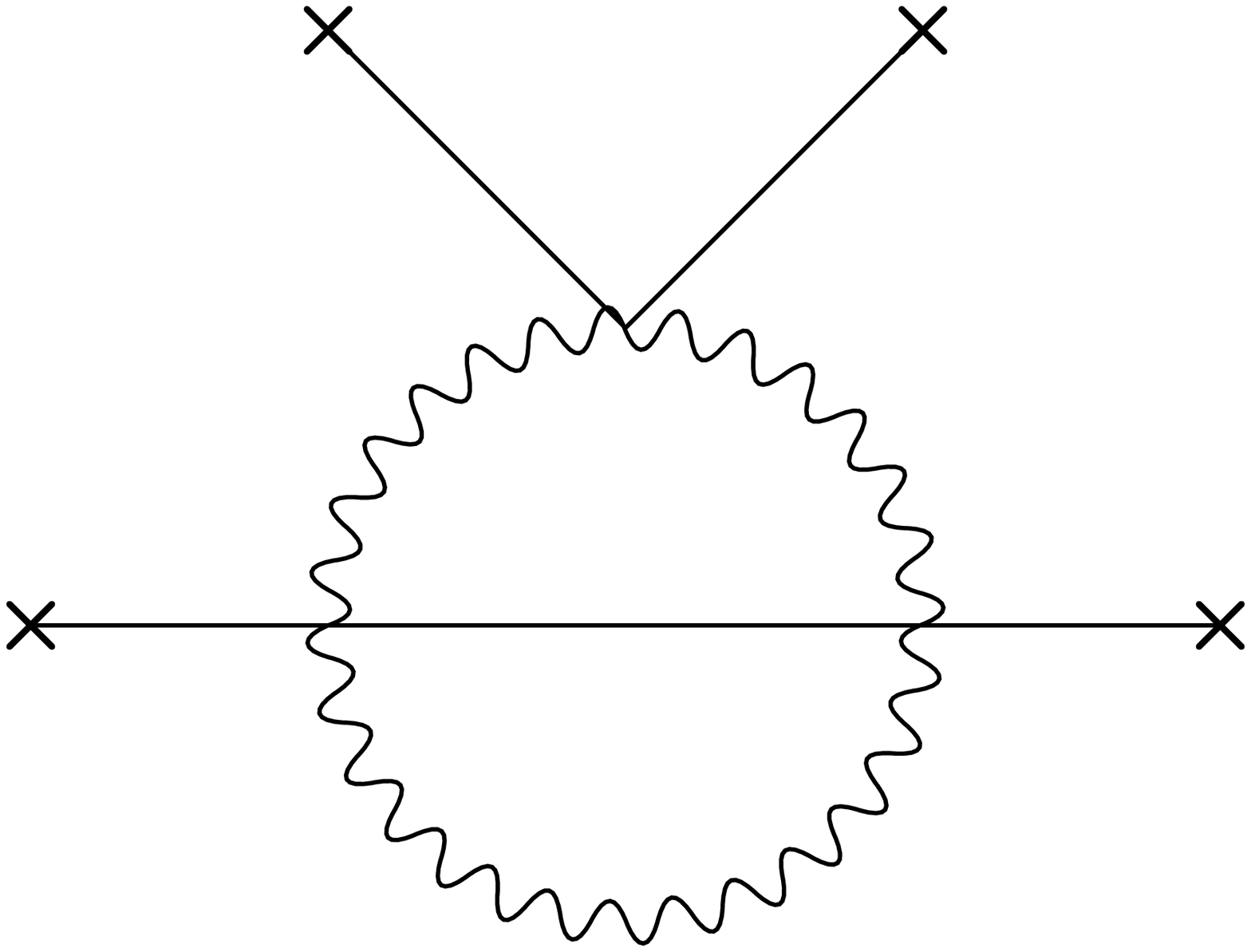,width=2cm}}\hspace{.4cm}
&=&0\label{fermigraph}
\eea
One may check that the fermion graphs\r{fermigraph} vanish, for the $K^{ab}$
pieces this cancellation is graph by graph as predicted above but for
 the pieces
with an $\epsilon(s-t)$ the four independent graphs (after considering all
combinations of dots) conspire to cancel and the relevant integrals
are given in table 1. Also we still need to use Riemann normal
coordinates to give a
covariant expression for\r{noncovariant}. Using
$\Gamma^\mu_{\alpha\beta}(O)=0=\omega_{\mu ab}(O)$,
$\partial_{\mu}\Gamma^\nu_{\alpha\beta}(O)=-(1/3)R^\nu\0_{(\alpha\beta)\mu}$
and $\partial_\mu\omega_{\nu ab}(O)=(1/2)R_{\mu\nu ab}$ we find
\be
\Box \partial_{[\mu}\omega_{\nu ]ab}=D^2R_{\mu\nu ab}+
\frac{1}{3}R^\gamma\0_{[\mu}R_{\nu]\gamma ab}
+\frac{1}{2}R_a\0^{e\alpha}\0_{[\mu}R_{\nu]\alpha eb}.
\ee
Hence we have for the two loop ($O(\hbar)$) contribution to ${\cal A}$
\be
{\cal A}_1=\int\bmu\epsilon^{abcd}
\left\{\begin{array}{c}
8D^2R_{\mu\nu ab}R^{\mu\nu}\0_{cd}-
8R_{\mu\nu}R^{\mu\alpha}\0_{ab}R^\nu\0_{\alpha cd}\\
-24R^\alpha\0_{\mu a}\0^eR_{\alpha\nu eb}R^{\mu\nu}\0_{cd}
-8R^{\mu(\nu\sigma)\rho}R_{\mu\nu ab}R_{\rho\sigma cd}\\
+4D_\alpha R^{\alpha\mu}\0 _{ab}D_\beta R^\beta\0_{\mu cd}
-5RR_{\mu\nu ab}R^{\mu\nu}\! \ _{cd}
\end{array}\right\},\label{zero}
\ee
where $\bmu\equiv d^4y\sqrt{g(y)}/\left( (2\pi i)^2 60\cdot 384\right)$.
It remains now only to show that the set of invariants
built from three Riemann tensors~\cite{japan}
in\r{zero} vanishes.
To this end one needs only the usual symmetries and
Bianchi identities for the Riemann tensor
\be
R_{\mu\nu\rho\sigma}=-R_{\nu\mu\rho\sigma}=-R_{\mu\nu\sigma\rho}=
R_{\rho\sigma\mu\nu};\ \
R_{\mu(\nu\rho\sigma)}=0=D_{(\alpha}R_{\mu\nu)\rho\sigma}
\ee
and the identity
\be
\delta^\mu_\nu\epsilon^{abcd}=
\delta^a_\nu\epsilon^{\mu bcd}+
\delta^b_\nu\epsilon^{a\mu cd}+
\delta^c_\nu\epsilon^{ab\mu d}+
\delta^d_\nu\epsilon^{abc\mu}.\label{pussycat}
\ee
Let us give some details. Rewrite the fourth term in\r{zero}
using $R^{\mu(\nu\sigma)\rho}R_{\mu\nu ab}=-(3/2)R^{\mu\nu\rho\sigma}
R_{\mu\nu ab}$ and apply\r{pussycat} twice to $\epsilon^{abcd}R=
R^{\rho\lambda}\0_{\sigma\eta}\delta^\sigma_\rho\delta^\eta_\lambda
\epsilon^{abcd}$ so that the sixth term in\r{zero}
becomes
\be
\int\bmu\epsilon^{abcd}RR^{\mu\nu}\0_{ab}R_{\mu\nu cd}
=\int\bmu\epsilon^{abcd}(2R^{\mu\nu\rho\sigma}R_{\mu\nu ab}R_{\rho\sigma cd}
-8R^\alpha\0_{\mu a}\0^eR_{\alpha eb\nu}R^{\mu\nu}\0_{cd}).
\ee
In a similar fashion one can rewrite the second term in\r{zero} as
\be
\int\bmu\epsilon^{abcd}R_{\mu\nu}R^{\mu\alpha}\0_{ab}R^\nu\0_{\alpha cd}
=\int\bmu\epsilon^{abcd}(-2R^\alpha\0_{\mu a}\0^eR_{\alpha b\nu e}
R^{\mu\nu}\0_{cd}+2R^\alpha\0_{\mu a}\0^eR_{\alpha eb\nu}R^{\mu\nu}\0_{cd}).
\ee
For the fifth term in\r{zero}, use the Bianchi identity on the indices
$\alpha ab$ and $\beta cd$, so that integrating by parts and using the
antisymmetry of $\epsilon^{abcd}$ one gets a commutator $[D_a,D_b]$
which may be expressed as curvatures whereby
\bea
\int\bmu\epsilon^{abcd}D_\alpha R^{\alpha\mu}\0 _{ab}
D_\beta R^\beta\0_{\mu cd}
&=&2\int\bmu\epsilon^{abcd}R^\mu_cR^\nu_dR_{ab\mu\nu}\nonumber\\
&=&\int\bmu\epsilon^{abcd}(-\frac{1}{2}R^{\mu\nu\rho\sigma}R_{\mu\nu ab}
R_{\rho\sigma cd}-2R^\alpha\0_{\mu a}\0^eR_{\alpha b\nu e}R^{\mu\nu}\0_{cd}
\nonumber\\
& & \hspace{1.6cm}+4R^\alpha\0_{\mu a}\0^eR_{\alpha eb\nu}R^{\mu\nu}\0_{cd}).
\eea
where the last line was obtained by using\r{pussycat}.
In a similar fashion the first term of\r{zero} may be expressed in terms
of curvatures as
\be
\int\bmu\epsilon^{abcd}D^2R_{\mu\nu ab}R^{\mu\nu}\0_{cd}
=\int\bmu\epsilon^{abcd}(4R^\alpha\0_{\mu a}\0^eR_{\alpha\nu eb}
R_{\mu\nu}\0_{cd}-4R^\alpha\0_{\mu a}\0^eR_{\alpha eb\nu}R_{\mu\nu}\0_{cd}).
\ee
Orchestrating the above manipulations, one finds
\be
{\cal A}_1=8\int\bmu\epsilon^{abcd}R^\alpha\0_{\mu a}\0^e
(R_{\alpha\nu eb}+R_{\alpha eb\nu}+R_{\alpha b\nu e})R^{\mu\nu}\0_{cd},
\ee
which clearly vanishes. This concludes our two loop demonstration
of the $\beta$-independence of the anomaly ${\cal A}$.

\section{Acknowledgements}
I would like to thank Peter van Nieuwenhuizen for suggesting this problem.
Further thanks are due to Kostas Skenderis for reading the manuscript.

\end{document}